\documentclass[aps,prx,twocolumn,superscriptaddress,nofootinbib,notitlepage,longbibliography]{revtex4-1}
\usepackage{graphicx}
\usepackage{amsmath}
\usepackage{amstext}
\usepackage{amssymb}
\usepackage{xfrac}
\usepackage[colorlinks,citecolor=blue]{hyperref}
\usepackage{graphicx}
\usepackage{amsmath}
\usepackage{amstext}
\usepackage{amssymb}
\usepackage{amsfonts}
\usepackage{longtable,booktabs}
\usepackage{hyperref}\usepackage{url}%Turn on when output to pdf file
\usepackage{subfigure}%
\usepackage{dsfont}
\usepackage{booktabs}
\usepackage{amsbsy}
\usepackage{dcolumn}
\usepackage{amsthm}
\usepackage{bm}
\usepackage{esint}
\usepackage{multirow}
\usepackage{hyperref}
\hypersetup{
    colorlinks=true,
    linkcolor=blue,
    filecolor=magenta,
    urlcolor=blue,
}
\usepackage{cleveref}

\usepackage{mathrsfs}
\usepackage{amsfonts}
\usepackage{amsbsy}
\usepackage{dcolumn}
\usepackage{bm}
\usepackage{multirow}
\usepackage{color}

\newcommand{\di}{\mathrm{d}}

\newcommand{\comments}[1]{}

  \usepackage{extarrows}
\usepackage{datetime}

\usepackage{comment}
\usepackage[super]{nth}
 
\begin{document}

\title{Fractonic Superfluids}
\author{Jian-Keng Yuan}\affiliation{School of Physics, Sun Yat-sen University, Guangzhou, 510275,
China}

\author{Shuai A. Chen}\email{s-chen16@mails.tsinghua.edu.cn}\affiliation{Institute for Advanced Study, Tsinghua University, Beijing,
100084, China}

\author{Peng Ye}
\email{yepeng5@mail.sysu.edu.cn}\affiliation{School of Physics, Sun Yat-sen University, Guangzhou, 510275,
China}
\date{\today}

\begin{abstract}
We propose a superfluid phase of ``many-fracton system''  in which charge and total dipole moments are conserved quantities. In this work, both microscopic model and long-wavelength effective theory are analyzed.   We start with a second quantized microscopic model and formulate the coherent-state path-integral representation. With repulsive interactions and positive chemical potential, we calculate various properties of the resulting superfluid state  and make comparison with a conventional superfluid. We deduce a highly nonlinear Euler-Lagrange equation as well as two Noether currents. We also formulate time-dependent Gross-Pitaevskii-type equations that govern hydrodynamical behaviors. We study the classical ground state wavefunction, the associated off-diagonal long range order (ODLRO), supercurrents, critical current, and unconventional topological vortices. At length scale much larger than coherence length $\xi_{\mathrm{coh}}$, we derive the effective theory of our microscopic model. Based on the effective theory, we analyze gapless Goldstone modes and specific heat capacity at low temperatures  as well as the fate of  ODLRO against quantum fluctuations. Several future directions, e.g., numerical analysis of Gross-Pitaevskii equations, fermionic fractons, fractonic superconductors, and cold-atom experimental realization, are discussed.
\end{abstract}

\maketitle

 \tableofcontents

\section{Introduction}
 \renewcommand\arraystretch{1.6}
\begin{table*}[tbp]
\caption{Comparison between conventional and fractonic superfuild phases. For simplicity, in this table, only isotropic case $K_{ij}=\frac{1}{2}\kappa$ ($\kappa>0$) of the microscopic model \eqref{Ham} is taken into account. 
 }
\label{Tab_key_result}
\begin{tabular}{p{4.2cm}<{\raggedright}p{5.5cm}<{\raggedright}p{5.5cm}<{\raggedright}}
\specialrule{0.1em}{0pt}{1.2pt}
%\hline
                                          &  {Conventional superfluid}   & {Fractonic superfluid} \\ \hline
% \specialrule{0em}{1.2pt}{1.2pt}
Order parameter $\langle \hat{\Phi}(x)\rangle$ &   $\sqrt{\rho_0}e^{i\theta_0 }$   & $\sqrt{\rho_0}e^{i(\theta_0+\sum_i\beta_i x^i) }$    \\
\multirow{2}*{Noether current}  &  \multirow{2}*{charge current $J=\rho_0^{}\mathbf{\nabla}\theta$ }   &
 charge current $J$ \eqref{U(1)-current},  dipole currents $\mathcal{D}^{(a)}$ \eqref{rank-1-current-Ja} \\
 %\specialrule{0em}{1.2pt}{1.2pt} 
Plane-wave dispersion   &      dispersive                                                          & dispersionless                                                        \\ 
Ground state                & $\exp [\!\int d^d x\, \sqrt{\rho _{0}}e^{i\theta_0 }\hat{\Phi}^{\dag }\!( \mathbf{x})] |0\rangle $                                                                                                     &\begin{small}$\exp [%
\int\! d^d x\, \sqrt{\rho _{0}}e^{i( \theta_0 +\sum_{i}^{d}\beta
_{i}x^{i}) }\hat{\Phi}^{\dag }\!( \mathbf{x}) ]
|0\rangle $ \end{small} \\
Topological number & $\ell=\oint_{C}\mathbf{v}\cdot d\mathbf{r}$ & $\ell =\oint_{C}\mathbf{U}\cdot d\mathbf{r} $ \\ 
Supercurrent               & charge current                                        & many-body current $\mathbf{\Xi}$ in Eq.~\eqref{CurXi}\\ 
Critical current            & $|J|_\mathrm{max}=\frac{2\sqrt{6}}{9} \sqrt{\frac{\mu^3}{g^2\kappa}}$                                                             & $(\mathbf{\Xi}_s)_\mathrm{max}=\frac{3\sqrt{3\kappa}\mu^2}{16\sqrt{g^3}}$ in Eq.~\eqref{DipileCurrentcrit} \\ 
Coherence length  $\xi _{\mathrm{coh}} $ &    $2\protect\pi\protect\sqrt{%
\protect\kappa /(4\protect\rho _{0}^{{}}g)}$  &  $2\protect\pi \sqrt[4]{\protect\kappa /4g}$  in Eq.~\eqref{cohlengtiso} \\
Goldstone mode       & $\omega \propto \left\vert \mathbf{k}\right\vert $ & $\omega \propto\left\vert \mathbf{k}\right\vert ^{2}$ \\ 
Stable dimension at $T=0$ & $d> 1$                                              & $d>2$ \\ 
 \specialrule{0.1em}{2pt}{0pt}
\end{tabular}%
\end{table*}
%EndExpansion

Liquid Helium-4 \cite{becbook,chaikin2000principles} is a typical quantum
many-boson system described by a Ginzburg-Landau theory. With interactions between bosons, superfluidity is established with formation of an off-diagonal long range order (ODLRO) \cite{YangODLRO} and emergence of  gapless Goldstone modes. 
Vortex configurations,
 which tend to eliminate  ODLRO, is topologically
characterized by the winding number of the circulating supercurrent.
Superfluid is also a simple demonstration on Mermin-Wagner (MW) theorem which
states that continuous symmetry cannot be spontaneously broken at any finite
temperatures ($T$) in one dimensional (1D) and 2D systems. At  zero temperature $T=0$, true
ODLRO is unstable against quantum fluctuations unless the spatial dimension is no less than two. 
Experimentally, achievements have been made on a variety of physical properties of superfluidity; meanwhile, superfluids serve as a platform for different fields, e.g. condensed matter, nuclear physics and high energy physics \cite{RevModPhys73307,RevModPhys80885,RevModPFermigas,2007AdPhy56243L,2009NJPh11e5049C,RevModPhys831523,2014RvMP86153G,Celi_2016,PhysRevB.93.115136,bti2,YeGu2015,yp18prl,yp18prl,WenSuperf}. Especially,  one may   consider a \emph{symmetric} phase formed by condensing  symmetry defects  in a superfluid or more general symmetry-breaking phases. By delicately designing  degrees of freedom on symmetry defects, one may construct symmetry-protected topological phases (SPT) \cite{Chen:2014aa,PhysRevB.93.115136,YeGu2015,bti2,yp18prl}.

In this paper, we propose an unconventional superfluid phase:
\emph{fractonic superfluid}, which was, surprisingly, motivated from seemingly uncorrelated  line of
thinking---strongly correlated topological phases of matter. Recently, there
is an ongoing focus issue---fracton topological order \cite{Chamon05,Vijay2015,Vijay2016a} that supports topological excitations with restricted
mobility. In contrast to the more ``familiar'' topological order  such
as the fractional quantum Hall effect, if one tries to move a fracton---a
point-like immobile excitation, additional fractons have to be created
nearby simultaneously. In other words, fractons are totally immobile. 
Tremendous progress has been made and vastly different research areas have
been unexpectedly connected in the context of fractons, such as glassy
dynamics, foliation theory, elasticity, dipole algebra, higher-rank global
symmetry, many-body localization, stabilizer codes, duality, gravity,
quantum spin liquid, and higher-rank gauge theory, see, e.g., review \cite{fracton_review1} and Refs.~\cite%
{Vijay2016a,Vijay2015,Vijay2016,Prem2017,Chamon05,Vijay2015,Shirley2019,Ma2017,Haah2011,Bulmash2019,Prem2019,Bulmash2018,Tian2018,You2018,Ma2018,Slagle2017,Halasz2017,Tian2019,Shirley2019b,Shirley2018a,Slagle2019a,Shirley2018,Prem2017,Prem2018,Pai2019,Pai2019a,Sala2019,Kumar2018,Pretko2018,Pretko2017,ye19a,Ma2018,Pretko2017a,Radzihovsky2019,Dua2019,PhysRevLett.122.076403,haahthesis,PhysRevX.9.031035,2019arXiv190411530Y,2019arXiv190408424S,2019arXiv190913879W,pretko18string,pretko18localization,PhysRevB.100.125150,PhysRevB.99.245135,PhysRevB.97.144106,PhysRevB.99.155118,MaHigherRankDQC, 2019arXiv191101804W,2019arXiv191213485W,2019arXiv191213485W,2020arXiv200503015D}.

While fractons are originally defined as point-like excitations, one may
also consider a \emph{many-fracton system}---a quantum many-body system directly made of
fractons. Suppose fractons are bosonic and simply represented by a scalar
field $\phi$, one may ask: what kind of minimal microscopic quantum models    can capture
the property of immobility? Ref.~\cite{Pretko2018} recently  proposed a non-Gaussian field theory by requiring that both total charge and
total dipole moments be conserved, where the time-derivative is \nth{2} order and the momentum-dependent term involves $\phi$ of at least \nth{4} order. This enhanced symmetry elegantly   enforces the mobility restriction of single particles.

Alternatively,   in this paper we  consider a 
minimal  model given by a second quantized microscopic Hamiltonian $\mathcal{H}$ that respects aforementioned symmetries.  Then, the coherent-state path integral quantization sends $%
\mathcal{H}$ to $\mathcal{L}= i\phi^*\partial_t \phi-\mathcal{H}$ after a Wick rotation.
 It should be noted that, the first-order derivative with respect to time in the Lagrangian $\mathcal{L}$ is very subtle. With this first-order derivative, one may legitimately interpret $\phi^*\phi$ as the
particle number density, which is a common situation in non-relativistic microscopic models in condensed matter physics 
and cold-atom. Starting from this Lagrangian, we consider a weak repulsive interaction in a grand canonical ensemble with positive chemical potential. The Euler-Lagrange  equation of this theory is highly non-linear, which is expected by observing that   the
Lagrangian as a functional of $%
\phi,\partial\phi$, and $\partial\partial\phi$ is intrinsically non-Gaussian. On the other hand, the Noether
currents associated to the two conserved quantities are derived: charge
current and dipole current.  Using a hydrodynamic approach\cite{becbook,chaikin2000principles}, we reformulate the Euler-Lagrange equation  as hydrodynamic equations which help understand superfluidity.  

We start with the normal state at $T=0$ with a  negative chemical potential. When chemical potential is turned to positive value, the energy functional drops down to minima when $\phi$ belongs to plane-wave configurations, in
contrast to conventional superfluid where $\phi$ of minima is exactly constant
everywhere, i.e., momentum $\mathbf{k}=0$. This class of configurations with lowest energy constitutes the
classical ground state manifold of fractonic superfluid, and the corresponding 
time-dependent Gross-Pitaevskii equations can be obtained in the presence of such exotic
boson condensate. 
In a conventional superfluid, the charge (or more precisely, particle number) current serves as   supercurrent. It can flow dissipationlessly as along as the current strength is below a critical value. The closed line integral of the supercurrent is topological in a sense that the numeric value of the integral   only depends on how many vortices are enclosed by the closed line, resulting in a quantized value.  Nevertheless, instead of charge current, in a fractonic superfluid,  we  have to identify a many-body current $\mathbf{\Xi}$ as a supercurrent such that it is topological and can flow dissipationlessly. In fact, both closed line integrals of  the two Noether currents mentioned above (charge and dipole   current) turn out   to be not topological. The corresponding
quantized number monitored by $\oint \mathbf{\Xi}$ represents unconventional topological vortices that are
expected to proliferate at critical points. Such a vortex shows interesting features and its dynamical behaviours deserve further investigation. 

The Goldstone
bosons associated to spontaneous broken symmetries in the many-fracton system are analyzed, whose
dispersion relations give rise to exotic temperature-dependence of specific
heat capacity $c_v$ as long as ODLRO is assumed.  Since quantum fluctuations are not treated seriously,  ODLRO
of classical ground states is  self-consistently established, regardless of dimensions. For this purpose, one can integrate out massive amplitude fluctuations, resulting in an effective field theory for phase fluctuations or the gapless Goldstone bosons based on our microscopic many-fracton model. In the so-called ``isotropic case'', the effective theory respects the Lifshitz spacetime symmetry and relates to nonrelativistic
gravity studied before \cite{Hoava2009, Hoava2009a, Xu2010}.  Once quantum fluctuations are taken into
account, Goldstone bosons and ODLRO are ultimately unstable in 1D and 2D. In 1D, the
correlation at long distance decays exponentially, which indicates a spectral gap is formed;   in 2D, it decays in
a power law.  Compared to the conventional superfluid phase, all these dimension-dependence properties of ODLRO arise as a result of highly non-Gaussianality of many-fracton systems.  A summary of comparison is given in Table~\ref{Tab_key_result}.

This paper is organized as follows. In Sec.~\ref{Sec:Hrs}, we introduce a microscopic Hamiltonian in Eq.~(\ref{Ham}) and it conserves total dipole moments as well as a charge. We derive the Euler-Lagrange equation and Noether currents. A Gross-Pitaevskii-type equation is also formulated to govern hydrodynamic behavior. Sec.~\ref{Sec:fracSF_classical} starts with a Mexican-hat potential to determine a fractonic superfluid phase in any spatial dimensions with the groundstate wavefunctions in Eq.~(\ref{gs_sf}) from a hydrodynamic method.  From the many-body current $\mathbf{\Xi}$ in Eq.~\eqref{CurXi}  that appears as the supercurrent, we define a new vector field  $\mathbf{U}$ in Eq.~(\ref{vorticity-u}) whose vorticity turns out to be topological. In Sec.~\ref{Sec:FS_QFT}, we concentrate on quantum fluctuations or gapless Goldstone modes. With an effective theory for Goldtone modes, we calculate correlators of order parameters, and   we give a temperature-dependence of specific heat capacity.  Lastly, Sec.~\ref{Sec:Conclusion} summarizes the main results and puts forward  further perspectives. A general many-fracton model is discussed in appendix at the end of the paper.

\section{Microscopic model}

\label{Sec:Hrs}
In this section, we introduce a microscopic model and derive the
Euler-Lagrange equation and the Noether currents, from which we recognize a Gross-Pitaevskii equation. All effective theory analysis in the remaining sections can be traced back to their microscopic origin   introduced below.

\subsection{Model Hamiltonian}

In a non-relativistic field theory, a single particle fails to propagate and may be eventually localized  if its effective mass  $M_{\mathrm{eff}} $ is too large. In other words, the usual kinetic term $\frac{1}{2M_{\mathrm{eff}}}\Phi^\dagger(-\nabla^2)\Phi $  vanishes. Nevertheless, it will be seen clear that  
mobility of bound state excitations can be independent on single particle mobility. Let us consider a non-quadratic Hamiltonian. One realization is a model $H=\int 
\mathrm{d}^{d}x\mathcal{H}(\hat{\Phi}^{\dag },\hat{\Phi})$ in $d$- (spatial) dimensional  manifold $M$, where $\mathcal{H}$ reads 
\begin{align}
\mathcal{H} =&\!\!\sum_{i,j}^{d}\!K_{ij}\! ( \hat{\Phi}^{\dag
}\partial _{i}\partial _{j}\hat{\Phi}^{\dag }-\partial _{i}\hat{\Phi}^{\dag
}\partial _{j}\hat{\Phi}^{\dag } ) \! ( \hat{\Phi}\partial
_{i}\partial _{j}\hat{\Phi}-\partial _{i}\hat{\Phi}\partial _{j}\hat{\Phi}%
 )   \notag \\
&+\sum_{i}^{d}G_{i}\!:\left( \partial _{i}\hat{\rho}\right)
^{2}:+V\! ( \hat{\Phi}^{\dag },\hat{\Phi} ) ,  \label{Ham}
\end{align}%
where $::$ denotes the normal ordering. $\hat{\Phi}^{\dag }\! ( \mathbf{x}%
 ) $ and $\hat{\Phi}( \mathbf{x} ) $ are bosonic creation
and annihilation operators and $\hat{\rho} ( \mathbf{x} ) =\hat{\Phi%
}^{\dag }\! ( \mathbf{x} ) \hat{\Phi} ( \mathbf{x} ) $ is
a density operator. These operators satisfy the standard communication relations 
\begin{align}
 &[ \hat{\Phi} ( \mathbf{x} ) ,\hat{\Phi}^{\dag } ( \mathbf{y} )  ] =\delta ^{d} ( \mathbf{x}-\mathbf{y} ), \label{com1}\\ 
 &[ \hat{\Phi} ( \mathbf{x} ) ,\hat{\rho} ( \mathbf{y} )  ] =\hat{\Phi} ( \mathbf{x} )\delta^d(\mathbf{x}-\mathbf{y}),  \label{com2}\\
 &[ \hat{\Phi}^{\dag }( \mathbf{x} ) ,\hat{\rho} ( \mathbf{y} )  ] =-\hat{\Phi}^{\dag }( \mathbf{x} ) \delta^d(\mathbf{x}-\mathbf{y})  \label{com3},\,
\end{align}
where $\mathbf{x}$ is a spatial coordinate and we neglect it in the following for symbol convenience. The potential  $V ( \hat{\Phi}^{\dag },\hat{\Phi})$ incorporates interactions that do not involve any spatial gradient. We  mainly
consider the simplest form of $V ( \hat{\Phi}^{\dag },\hat{\Phi} )$
\begin{equation}
V ( \hat{\Phi}^{\dag },\hat{\Phi} ) =-\mu \hat{\Phi}^{\dag }\hat{%
\Phi}+\frac{g}{2}\hat{\Phi}^{\dag }\hat{\Phi}^{\dag }\hat{\Phi}\hat{\Phi}\,,
\label{v-sq}
\end{equation}%
where $\mu $ is the chemical potential and $g>0$ describes onsite repulsive
interaction. \emph{Hereafter, no Einstein summation rule is assumed}. The coupling
constants $K_{ij}>0$ and $G_{i}\geq 0$ ensure a lower bound for a physically acceptable Hamiltonian $%
\mathcal{H}$. The $G_{i}$-term is also a potential term.  
 Besides, no rotational invariance is assumed since
anisotropy of $K_{ij}$ is allowed. 
The Hamiltonian in Eq.~\eqref{Ham} is possible to be realized in cold atomic gas subjected to an optical lattice by simulating the correlated hopping of two bosons \cite{ExperimentFisher2005,2008PhRvA77b3603P}. Hamiltonian $\mathcal{H}$ obeys a conservation law of total dipole moments as well as a
global $U\left( 1\right) $ symmetry. So we have two types of conserved
quantities. One is the global $U( 1) $ charge $\hat{Q}=\int \!\mathrm{d}^{d}\mathbf{x}\,\hat{\rho}$  and the others are the total dipole moments of $d$ components
$\hat{Q}^{(a)}=\int \!\mathrm{d}^{d}\mathbf{x}\,\hat{\rho}x^{a}\,, \,a=1,2,\cdots,d\,,$
where the integral is over the $d$-dimensional spatial manifold $M$. We denote a group
generated by both $\hat{Q}$ and $\hat{Q}^{(a)}$ $\left( a=1,2,\cdots ,d\right) $ as $%
\mathcal{G}$\cite{Seiberg2019arXiv1909}. We denote a subgroup with a single generator $Q^{(a)}$ as $%
U(1)^{a}$. An element $U=\exp \left[ -i\left( \lambda \hat{Q}+\sum_{a}\lambda
_{a}\hat{Q}^{\left( a\right) }\right) \right] $ in $\mathcal{G}$ leads to a
transformation of the field $\phi $ as $\phi ^{\prime }=U\phi U^\dagger=\phi e^{i\left(
\lambda +\sum_{a}\lambda _{a}x^{a}\right) }$ with $d+1$ real parameters $%
\lambda$ and $\lambda_{a}$ $\left( a=1,\cdots ,d\right) $. The group $%
\mathcal{G}$ is not an internal symmetry because $\hat{Q}^{(a)}$ does
not commute with the translational or rotational symmetry. 
 
By performing the  coherent-state path integral quantization, we can
construct a partition function $\mathcal{Z}=\int \mathcal{D}\phi \mathcal{D}\phi^* e^{i\int d^d\mathbf{x} dt\mathcal{L}}$ with Lagrangian $\mathcal{L}$ from $\mathcal{H}$ in Eq.~(\ref{Ham})
as 
\begin{equation}
\mathcal{L}=i\phi ^{\ast }\partial _{t}\phi -\mathcal{H}\left( \phi ^{\ast
},\phi \right) ,  \label{Lagrangian}
\end{equation}%
where $\phi \left( \mathbf{x},t\right) $ is the eigenvalue of annihilation
operator $\hat{\Phi}\!\left( \mathbf{x}\right) $ on a coherent state
\begin{align}
\hat{\Phi}\left( \mathbf{x}\right) |\phi \left( \mathbf{x},t\right) \rangle =\phi
\left( \mathbf{x},t\right) |\phi \left( \mathbf{x},t\right) \rangle
\end{align}
and $\phi ^{\ast }\left( \mathbf{x},t\right) $ is its complex conjugate. It should be noted that a Wick
rotation has been applied from imaginary time to real time,  which is convenient for the physics of zero temperature. 
The subtle ``first-order time derivative'' term in Eq.~(%
\ref{Lagrangian}) is ultimately determined by  Hamiltonian \eqref{Ham} and commutation  relations listed in Eqs.~\eqref{com1}, \eqref{com2} and \eqref{com3}, which can be verified by performing the standard canonical quantization. 
\textit{We regard Eqs.~(\ref{Ham}) and (\ref{Lagrangian}) as the microscopic model of fractonic superfluids.} 
 
%%%%%%%%%%%%%%%%%%%%
% 

%
 
\subsection{Euler-Lagrange equation and Noether theorem\label{Sec:NoetherThm}}

The Noether theorem states that a classical action that respects a continuous symmetry is associated to a conserved
charge. A continuity equation can be deduced from the action. Below we will derive the Euler-Lagrange equation as well as Noether currents from the Noether theorem. Due to the non-Gaussian nature of the microscopic model,  the derivation below will demonstrate several exotic features that do not appear  in usual Gaussian models.

First of all, we derive the Euler-Lagrange equation. Generally the stationary condition of an action $S=\int \mathrm{d}^{d}\mathbf{x}\mathrm{d}t\mathcal{L}\left[ \partial _{t}\phi ,\partial _{i}\phi ,\partial
_{i}\partial _{j}\phi ,\phi \right] $ meets the Euler-Lagrange equation.
Here for the notational convenience, we do not  explicitly show the dependence on $\phi ^{\ast
} $ and its derivative terms in $\mathcal{L}$. A variation $%
\delta \phi $ leads to $\delta S$ 
\begin{align}
\delta S &=\int \mathrm{d}^{d}\mathbf{x}\mathrm{d}t\delta \mathcal{L}\left[ \partial
_{t}\phi ,\partial _{i}\phi ,\partial _{i}\partial _{j}\phi ,\phi \right] 
\notag \\
&=\int \mathrm{d}^{d}\mathbf{x}\mathrm{d}t\left[ \frac{\delta \mathcal{L}}{\delta
\phi }-\partial _{t}\frac{\delta \mathcal{L}}{\delta \partial _{t}\phi }%
-\sum_{i}^{d}\partial _{i}\frac{\delta \mathcal{L}}{\delta \partial
_{i}\phi }\right.  \notag \\
&\left. +\sum_{i,j}^{d}\partial _{i}\partial _{j}\frac{\delta \mathcal{L%
}}{\delta \partial _{i}\partial _{j}\phi }\delta \phi \right] +\text{c.c}\,.
\label{deltaS}
\end{align}%
Here, the variation does not depend on space-time coordinates, $\delta
\partial _{i}\phi =\partial _{i}\delta \phi ,\delta \partial _{i}\partial
_{j}\phi =\partial _{i}\partial _{j}\delta \phi ,$ etc and c.c. means complex conjugate.

Up to a surface term, vanishing of Eq.~(\ref{deltaS}) requires the Euler-Lagrange   equation,%
\begin{align}
\partial _{t}\frac{\delta \mathcal{L}}{\delta \partial _{t}\phi } &=\frac{%
\delta \mathcal{L}}{\delta \phi }-\sum_{i}^{d}\partial _{i}\frac{\delta 
\mathcal{L}}{\delta \partial _{i}\phi }+\sum_{i,j}^{d}\partial
_{i}\partial _{j}\frac{\delta \mathcal{L}}{\delta \partial _{i}\partial
_{j}\phi } .
\label{abstract_Euler-Lagrange}
\end{align}%
One remark is that we take $\partial _{i}\partial _{j}\phi $ and $\partial
_{j}\partial _{i}\phi $ as different variables if $i\not=j$ during
variational processes,%
\begin{equation}
\frac{\delta \partial _{i}\partial _{j}\phi ( \mathbf{x},t) }{\delta
\partial _{m}\partial _{n}\phi (\mathbf{ x}^{\prime },t^{\prime }) }%
=\delta _{im}\delta _{jn}\delta (\mathbf{x}-\mathbf{x}^{\prime }) \delta (
t-t^{\prime }) .
\end{equation}

In sharp contrast to the usual Euler-Lagrange equations, there are three terms in   the r.h.s. of Eq.~(\ref{abstract_Euler-Lagrange}) where the third term arises from the non-Gaussionality. Plugging  Eq.~(\ref{Lagrangian}) into Eq.~(\ref{abstract_Euler-Lagrange}) renders %
\begin{align}
i\partial _{t}\phi &=\sum_{i,j}^{d}K_{ij}\partial _{i}\partial _{j}\left[
\phi ^{\ast }\left( \phi \partial _{i}\partial _{j}\phi -\partial _{i}\phi
\partial _{j}\phi \right) \right]  \notag \\
&+2K_{ij}\partial _{i}\left[ \partial _{j}\phi ^{\ast }\left( \phi \partial
_{i}\partial _{j}\phi -\partial _{i}\phi \partial _{j}\phi \right) \right] 
\notag \\
&+K_{ij}\partial _{i}\partial _{j}\phi ^{\ast }\left( \phi \partial
_{i}\partial _{j}\phi -\partial _{i}\phi \partial _{j}\phi \right)  \notag \\
&-\sum_{i}^{d}2G_{i} \partial _{i}^{2}\rho \,\phi -\mu \phi +g\rho \phi\,,
\label{EL}
\end{align}%
where we have considered  a Mexican-hat potential in Eq.~(\ref{Lagrangian}):
\begin{equation}
V ( \phi  ) =-\mu \left\vert \phi \right\vert ^{2}+\frac{g}{2}%
\left\vert \phi \right\vert ^{4}\,. \label{g-potential}
\end{equation}%
It is the path-integral representation of    the operator form $V ( \hat{\Phi}^{\dag },\hat{\Phi} ) $
in Eq.~(\ref{v-sq}). 

Brute-forcely solving Eq.~(\ref{EL}), both numerically and analytically, is not easy due to its high non-linearity. Nevertheless, one may   quickly verify the existence of  immobile fractons just by taking a plane-wave ansatz 
\begin{align}
\phi =\mathscr{N}\exp \left( i\omega t-i\mathbf{k}\cdot 
\mathbf{x}\right) .
\end{align}  
where $\mathscr{N}$ is a proper normalization factor. 
The flat dispersion relation $\omega =0$  indicates that a single particle is non-propagating. Such kinds of particles  with fully restricted mobility are dubbed   \emph{fracton} in the literature of fracton topological order.  

Now we are in a position to  calculate the Noether currents associated with the two conserved quantities. 
We consider on-shell variations where fields $\phi $ and $%
\phi ^{\ast }$ are constrained to satisfy Euler-Lagrange equations while the variations 
$\delta \phi $ and $\delta \phi ^{\ast }$ are arbitrary. A symmetry
transformation%
\begin{equation}
\phi \rightarrow \phi ^{\prime }=\phi +\delta \phi =\phi +\alpha F\left(
\phi \right)
\end{equation}%
has a parameter $\alpha $ that is independent of space-time coordinates and
keeps the Euler-Lagrange equation invariant while it does not involve changes in the
coordinates in any way. The Noether theorem states that even when $\alpha $
depends on coordinates $\alpha =\alpha( \mathbf{x},t) $, the variation
action $\delta S =\int \di^{d}\mathbf{x}\di t\mathcal{L}\left[ \partial _{t}\phi ^{\prime
},\partial _{i}\phi ^{\prime },\partial _{i}\partial _{j}\phi ^{\prime
},\phi \right] -\mathcal{L}\left[ \partial _{t}\phi ,\partial_i\phi, \partial_i\partial_j\phi,\phi \right]$ should also vanish. 
\begin{widetext}
\begin{align}
\delta S &=\!\int \!\di^{d}\mathbf{x}\di t\frac{\delta \mathcal{L}}{\delta \partial _{t}\phi }\delta
\partial _{t}\phi +\sum_{i}^{d}\frac{\delta \mathcal{L}}{\delta \partial
_{i}\phi }\delta \partial _{i}\phi +\sum_{i,j}^{d}\frac{\delta \mathcal{L}}{\delta \partial _{i}\partial
_{j}\phi }\delta \partial _{i}\partial _{j}\phi +\frac{\delta \mathcal{L}}{%
\delta \phi }\delta \phi  \notag \\
&=\!\int\! \di^{d}\mathbf{x}\di t(-\alpha )\partial _{t}\!\left( \frac{\delta \mathcal{L}}{%
\delta \partial _{t}\phi }F\right) -\sum_{i}^{d}\alpha \left[ \partial
_{i}\left( \frac{\delta \mathcal{L}}{\delta \partial _{i}\phi }F\right)
+\sum_{i,j}^{d}\partial _{i}\!\left( \frac{\delta \mathcal{L}}{
\delta \partial _{i}\partial _{j}\phi }\partial _{j}F + \frac{\delta \mathcal{L}}{
\delta \partial _{j}\partial _{i}\phi }\partial _{j}F\right )
-\sum_{i,j}^{d}\partial _{i}\partial _{j}\left( \frac{\delta \mathcal{L}}{%
\delta \partial _{i}\partial _{j}\phi }F\right) \right],
\end{align}%
\end{widetext}
where the Euler-Lagrange equation in Eq.~(\ref{abstract_Euler-Lagrange}) is applied. The variation $\delta S$ appears as an
integral over a total derivative 
\begin{equation}
\delta S =\alpha\int \mathrm{d}^{d}\mathbf{x}\mathrm{d}
t\left(\partial _{t}\rho +\sum_{i}^{d}\partial _{i}J_{i}\right)
\end{equation}%
We arrive at conserved charge $Q$ and current densities $J_i$ 
\begin{align}
Q &=-\int \mathrm{d}^{d}\mathbf{x}\frac{\delta \mathcal{L}}{\delta \partial _{t}\phi 
}F+\text{c.c}=\int \mathrm{d}^{d}x\rho  \,,\notag \\
J_{i} &=-\left( \frac{\delta \mathcal{L}}{\delta \partial _{i}\phi }%
-\sum_{j}^{d}\partial _{j}\frac{\delta \mathcal{L}}{\delta \partial
_{i}\partial _{j}\phi }\right) F-\sum_{j}^{d}\frac{\delta \mathcal{L}}{%
\delta \partial _{i}\partial _{j}\phi }\partial _{j}F+\text{c.c}
\end{align}%
and the conservation law 
\begin{equation}
\partial _{t}\rho +\sum_{i}^{d}\partial _{i}J_{i}=0\,.
\end{equation}
Back to our model in Eq.~(\ref{Lagrangian}), for a global $U\left( 1\right) $ symmetry, we take $F( \phi )
=i\phi $ and $F( \phi ) =ix^{a}\phi $ for $U( 1)
^{( a) }$ and we can obtain charge and current densities, 
\begin{align}
Q &=\int \!\mathrm{d}^{d}\mathbf{x}\,\phi ^{\ast }\phi  \label{U(1)-charge} \\ 
J_{i} &=i\sum_{j}^d K_{ij}\partial _{j}\left[ \phi ^{\ast 2}\left( \phi\partial _{i}\partial _{j}\phi -\partial _{i}\phi \partial _{j}\phi \right) -\text{c.c.}\right]  \label{U(1)-current} \\
Q^{\left( a\right) } &=\int \!\mathrm{d}^{d}\mathbf{x}\,x^{a}\phi ^{\ast }\phi
\label{rank-1-charge-Qa} \\
\mathcal{D} _{i}^{\left( a\right) } &=i\sum_{j}^{d}K_{ij}x^{a}\partial _{j} \left[ \phi ^{\ast 2}\left( \phi \partial _{i}\partial _{j}\phi -\partial_{i}\phi \partial _{j}\phi \right) -\text{c.c.}\right]  \notag \\
&-i\sum_j^{d}K_{ij}\delta _{a}^{j}\left[ \phi ^{\ast 2}\left( \phi\partial _{i}\partial _{j}\phi -\partial _{i}\phi \partial _{j}\phi \right) -\text{c.c.}\right] .  \label{rank-1-current-Ja}\end{align}
Therefore, we have two types of  spatial currents: $J_{i}$ and $\mathcal{D} _{i}^{\left( a\right)
} $. Nevertheless, they are not totally independent. The first term in current $\mathcal{D}
_{i}^{( a) }$ in Eq.~(\ref{rank-1-current-Ja}) that equals $x^{a}J_{i}$
comes from motions of each single particle at $\mathbf{x}$ with current 
$J_{i}$ and the extra term comes from the  pure effect during  many-body hopping processes. It motivates us to isolate the many-body current $\mathbf{\Xi}$
\begin{equation}
\Xi_{ia}=x^{a}J_i-\mathcal{D}^{(a)}_i .
\label{CurXi}
\end{equation}
The many-body current $\mathbf{\Xi}$ is symmetric under its index and  it has relation with charge current
$
J_i=\sum_{a=1}^d \partial_a \Xi_{ia}. 
$
This relation implies a generalized conversation law
$
\partial_t \rho+\partial_i\partial_a \Xi_{ia}=0 .
$
As we will see in Sec.~\ref{Sec:fracSF_classical} that the current $\mathbf{\Xi}$ plays a vital role.

\subsection{Time-dependent Gross-Pitaevskii-type equations}

\label{Sec:GPandhydro}

Below, we will deduce equations that govern hydrodynamic behaviors of the superfluid, which are summarized as a time-dependent Gross-Pitaevskii   equation set. 
We now rewrite  Eq.~(\ref{EL}) 
\begin{equation}
i\partial _{t}\phi =\hat{H}\phi ,  \label{Gpequ}
\end{equation}%
where $\hat{H}$ behaves as a single-particle Hamiltonian that reads%
\begin{align}
\hat{H} =&\sum_{i,j}^{d}K_{ij}\partial _{i}\partial _{j}\left[ \phi
^{\ast }\left( -\partial _{i}\phi \partial _{j}+\phi \partial _{i}\partial
_{j}\right) \right]  \notag \\
&+2K_{ij}\partial _{i}\left[ \partial _{j}\phi ^{\ast }\left( \phi \partial
_{i}\partial _{j}-\partial _{i}\phi \partial _{j}\right) \right]  \notag \\
&+K_{ij}\partial _{i}\partial _{j}\phi ^{\ast }\left( \phi \partial
_{i}\partial _{j}-\partial _{i}\phi \partial _{j}\right)  \notag \\
&-\sum_{i}^{d}2G_{i}\partial _{i}^{2}\rho -\mu +g\rho .
\end{align}%
Eq.~(\ref{Gpequ}) has a similar form as a time-dependent Gross-Pitaevskii
equation in a conventional superfluid phase where $g$ characterizes a
hardcore interaction. Differently the kinetic term is nonlinear due to
refinement from the symmetry group $\mathcal{G}$. One way to understand Eq.~(\ref{Gpequ}) is to
derive a hydrodynamic equation by decomposing $\phi =\sqrt{\rho }e^{i\theta
} $ where the real fields $\rho $ and $\theta $ are density and phase operators
respectively. So, the Gross-Pitaevskii equation is equivalent to two partial derivative
equations, 
\begin{align}
\!\!\!\!\frac{\partial \rho }{\partial t} \!= &2\sum_{i,j}^{d}K_{ij}\partial
_{i}\partial _{j}\left( \rho ^{2}\partial _{i}\partial _{j}\theta \right)\,,
\label{hydro-eq-rho} \\
\!\!\!\!\!\! \!\!\frac{\partial \theta }{\partial t} \!=&-\frac{1}{2\rho ^{3}}%
\sum_{i,j}^{d}K_{ij}\left[ (\partial _{i}\rho \partial _{j}\rho
)^{2}-2\rho \partial _{i}\rho \partial _{j}\rho \partial _{i}\partial
_{j}\rho \right]  \notag \\
&\!\!\!\!\!-\frac{1}{2\rho }\sum_{i,j}^{d}K_{ij}\left[ (\partial _{i}\partial
_{j}\rho )^{2}-\partial _{i}^{2}\rho \partial _{j}^{2}\rho \right]  \notag \\
&\!\!\!\!\!\!-\!\frac{1}{2}\!\sum_{i,j}^{d}\!\!K_{ij}\!\!\left[ 4\rho (\partial _{i}\partial
_{j}\theta )^{2}\!\!+\!\!\partial _{i}^{2}\partial _{j}^{2}\rho \right]\!
+\!\!\sum_{i}^{d}\!G_{i}\partial _{i}^{2}\rho \!+\!\mu \!-\!\!g\rho .\!\!\!\label{hydro-eq-theta}
\end{align}%
Eq.~(\ref{hydro-eq-rho}) is a continuity equation and the dynamics of $%
\theta $ is very complicated. The solution towards Eqs.~(\ref{hydro-eq-rho})
and (\ref{hydro-eq-theta}) resembles a fluid with conserved dipole moments.

The hydrodynamic velocity $v_{i}$ is defined as 
\begin{equation}
J_{i}=\rho v_{i} \, ,
\end{equation}%
where $\rho $ is the charge density. From $J_{i}$ in Eq.~(\ref{U(1)-current}%
), we find that 
\begin{equation}
v_{i}=-\sum_{j}^{d}2K_{ij}\left( 2\partial _{j}\rho \partial _{i}\partial
_{j}\theta +\rho \partial _{i}\partial _{j}^{2}\theta \right) \, .
\label{rank-0-velocity}
\end{equation}%
The ``velocity'' $v_{i}^{\left( a\right) }$ with a relation $\mathcal{D}
_{i}^{\left( a\right) }=\rho v_{i}^{\left( a\right) }$ can also be deduced
from Eq.~(\ref{rank-1-current-Ja}) as%
\begin{equation}
v_{i}^{\left( a\right) }=2K_{ia}\rho \partial _{i}\partial _{a}\theta
+v_{i}x^{a} \, . \label{rank-1-velocity}
\end{equation}%
  It's easy to extract
two continuity equations 
\begin{align}
\frac{\partial \rho }{\partial t}+\sum_{i}^{d}\partial _{i}\left( \rho
v_{i}\right) &=0 \, , \label{continueEq1} \\
\frac{\partial \rho ^{\left( a\right) }}{\partial t}+\sum_{i}^{d}\partial
_{i}\left( \rho v_{i}^{\left( a\right) }\right) &=0 \, . \label{continueEq2}
\end{align}%
Numerical simulations to Eqs. (\ref{hydro-eq-rho}) and (\ref{hydro-eq-theta}%
) may show interesting features, which can help us get insight into the GP
equation in Eq.~(\ref{Gpequ}), and it deserves future investigations. Before moving to next section, we should emphasize that all equations, currents and charges are not specified to a certain phase of the microscopic model. In the next section, we will focus on the superfluid phase.

\section{Fractonic superfluidity}
\label{Sec:fracSF_classical}
 We have discussed basic properties like Noether
currents and Gross-Pitaevskii equation in a many-fracton model in Eq.~\eqref{Ham}. The main feature is its
non-Gaussian form resulting from a dipole-moment conservation symmetry $%
\mathcal{G}$. In this section, we discuss in details the fractonic superfluidity arising from our microscopic model.

\subsection{ODLRO and order parameter}

Superfluidity can occur in a conventional bosonic system with a potential $%
V\left( \phi \right) $ in Eq.~(\ref{g-potential}). In this section, we
consider a superfluid phase in a fracton system in the microscopic model  (\ref{Ham}).

Classically, the energy density $\mathcal{E}$ for the steady system in Eq.~\eqref{Ham} has the form as 
\begin{align}
\mathcal{E} =&\sum_{i,j}^{d}K_{ij}\left\vert \phi \partial _{i}\partial _{j}\phi
-\partial _{i}\phi \partial _{j}\phi \right\vert ^{2}  \notag \\
&+\sum_{i}^{d}G_{i}(\partial _{i}\rho )^{2}-\mu \left\vert \phi
\right\vert ^{2}+\frac{g}{2}\left\vert \phi \right\vert ^{4}
\label{TotalE}
\end{align}%
and the field configuration $\phi$ at its minimum depends on the chemical potential. If $\mu <0$, the potential $V\left( \phi
\right) $ has a minimal value at $\rho =0$. It is a normal phase. If $\mu
>0 $, the potential $V( \phi ) $ reaches a minimal value at $%
\left\vert \phi \right\vert =\sqrt{\rho _{0}}\equiv \sqrt{\frac{\mu }{g}}$.
The vacuum now possesses a finite particle density and thus a large number of
degeneracies. In the \nth{2} quantization language, the groundstate manifold
can be represented with a creation operator $\hat{\Phi}^{\dag}$ along with
phase parameters $\theta_0 $ and $\beta _{i}$ $\left( i=1,\cdots ,d\right) $ 
\begin{equation}
|\text{GS}_{\beta _{i}}^{\theta_0 }\rangle =\bigotimes_{\mathbf x}|\text{GS}_{\beta
_{i}}^{\theta_0 }\rangle _{\mathbf x}\,,
\label{gs_sf}
\end{equation}%
where $|\text{GS}_{\beta _{i}}^{\theta_0 }\rangle_\mathbf{x}$ describes particles at position $\mathbf{x}$
\begin{equation}
|\text{GS}_{\beta _{i}}^{\theta_0 }\rangle _{\mathbf{x}}=\frac{1}{C}\exp \left[ \sqrt{%
\rho _{0}}e^{i\left( \theta_0 +\sum_{i}^{d}\beta _{i}x^{i}\right) }\Phi^{\dag
}\left( \mathbf{x}\right) \right] |0\rangle~,
\label{wfSingleParticle}
\end{equation}%
with $C=e^{\frac{1}{2}\rho _{0}}$ as the normalization factor.
For two such ground states $|\mathrm{GS%
}_{\beta _{i}}^{\theta_0 }\rangle $ and $|\mathrm{GS}_{\beta _{i}^{\prime
}}^{\theta_0 ^{\prime }}\rangle $ with $\Delta \theta_0 =\theta_0 -\theta_0 ^{\prime },\Delta \beta _{i}=\beta _{i}-\beta
_{i}^{\prime }$, from their inner product, 
\begin{align}
&\left\vert \langle \text{GS}_{\beta _{i}^{\prime }}^{\theta_0 ^{\prime }}|%
\text{GS}_{\beta _{i}}^{\theta_0 }\rangle\right\vert^2  \notag \\
=&\left\{ 
\begin{array}{cc}
\exp \left( -2\rho _{0}V\right)    & \mathrm{if}\,\Delta \beta _{i}\not =0 \quad    \exists  i \\ 
\exp \left[ -2\rho _{0}V\left( 1-\cos \Delta \theta_0 \right) \right]  & \mathrm{if }\,\Delta \beta _{i}=0 \quad  \forall i%
\end{array}%
\right.   \label{App:orth}
\end{align}%
we can conclude that they are orthogonal in the thermodynamic limit  $V\rightarrow \infty$ where $V$ is the volume of the spatial manifold $M$.
The ground state  in Eq.~(\ref{gs_sf}) comprises equal-weight
superposition over all possible numbers of particles that is modulated by a
phase factor and it 
characterizes  condensation of a macroscopically large number of particles at a state with momentum $\mathbf{k}=(\beta_1,\cdots,\beta_d)$ by observing $|\mathrm{GS}_{\beta _{i}}^{\theta_0 }\rangle=\exp[\sqrt{\rho_0}e^{i\theta_0}\hat{\Phi}^{\dag}(\mathbf{k})]|0\rangle$ where $ \hat{\Phi}^{\dag}(\mathbf{k})$ is the Fourier transformation of $\hat{\Phi}^{\dag}(\mathbf{x})$.  We call state $|\mathrm{GS}_{\beta _{i}}^{\theta_0 }\rangle$ in Eq.~(%
\ref{gs_sf}) as a \emph{fractonic superfluid} phase.  The most significant
feature of the state (\ref{gs_sf}) is the formation of an off-diagonal long range
order (ODLRO). If we calculate the correlation function in the classical
level 
\begin{equation} 
\!\!\!\!\!\!\!\!C\left( \mathbf{x}\right) =\langle \text{GS}_{\beta _{i}}^{\theta_0 }|\hat{%
\Phi}^{\dag }\!\left( \mathbf{x}\right) \hat{\Phi}\!\left( 0\right) |\text{GS%
}_{\beta _{i}}^{\theta_0 }\rangle  =\rho _{0}e^{-i\left( \sum_{i}^{d}\beta _{i}x^{i}\right) }\,
\label{correlation_cl}
\end{equation}%
whose amplitude doesn't decay  at  large distances. 
Equivalently, an order parameter  can be determined with finite expectation value on the ground state 
\begin{equation}
\langle \hat{\Phi}(\mathbf{x})\rangle=\langle \text{GS}_{\beta _{i}}^{\theta_0 }|\hat{\Phi}(\mathbf{x}) |\text{GS%
}_{\beta _{i}}^{\theta_0 }\rangle   =\sqrt{\rho_0}e^{i(\theta_0+\sum_i \beta_i x^i)}~.
\label{orderpara}
\end{equation}

\subsection{Supercurrent and its critical value}
\label{Sec:SuperCurr}

Currents can suppress superfluidity. In a conventional superfluid phase, frictionless charge current can exist if superfluidity is not totally destroyed. In other words, superfluidity can survive as long as the system has a finite order parameter or particle density $\rho_0$  for the minimal total energy. We can expect a critical current as the maximum one that a conventional superfluid phase can sustain. 
Similar discussion can be applied to a fractonic superfluid phase. Here we investigate the critical current in a fractonic superfluid phase in an isotropic case $K_{ij}=\frac{1}{2}\kappa$.

The groundstate in Eq.~\eqref{gs_sf} can be considered as the one that minimizes the  energy density $\mathcal{E}$ in Eq.~\eqref{TotalE} under a specified  boundary condition on the net phase change. For example, a wavefunction $|\mathrm{GS}^{\theta_0}_{\beta_i}\rangle$  minimizes  $\mathcal{E}$ in Eq.~\eqref{TotalE} under the  boundary conditions
%\begin{align}
  %\end{align}
\begin{subequations}
  \begin{equation}
    \label{BC-a}
      \Delta_i\theta=\beta_i L  \quad  i=1,\cdots,d
  \end{equation}
  \begin{equation}
    \label{BC-b}
   \Delta_i\partial_j\theta =0 \quad  i,j=1,\cdots,d  
  \end{equation}
\end{subequations}
where $\Delta_i\theta$ ($\Delta_i \partial_j\theta$) is the net difference of field  $\theta$ ($ \partial_j\theta$) along $x^i$-direction with $L$ as the system size. 
A possible suppressing factor is  equivalent to twisting the boundary conditions in Eqs.~\eqref{BC-a}  and \eqref{BC-b}. For example, the net change of $\partial_j\theta$ can be twisted to a finite value, which will invalidate the form of field $\theta=\theta_0+\sum_{i=1}^d \beta_i x^i$ in Eq.~\eqref{gs_sf}. 
Our aim is to determine the order parameter and currents under a twisted boundary condition. 
 Minimizing  the energy  density $\mathcal{E}$ in Eq.~\eqref{TotalE} requires uniformity of $|\nabla^2\theta(x)|$ and we denote $\nu=|\nabla^2\theta(x)|$.  Based on it, in a superfluid phase with $\mu>0$, $\mathcal{E}$ in Eq.~\eqref{TotalE} can be written as 
\begin{equation}
\mathcal{E}=\frac{1}{2}\kappa \rho^2 |\nu|^2-\mu \rho+\frac{g}{2}\rho^2\,.
\label{FreeEnergy}
\end{equation}
where $\rho$ is  particle density of condensate field or order parameter $\hat{\Phi}$ and we assume it uniform in space $\nabla\rho=0$.   Minimize $\mathcal{E}$ with respect to $\rho$, and we obtain the particle density 
\begin{equation}
\rho_0=\frac{\mu}{\kappa \nu^2+g} 
\end{equation}
 which is suppressed by $\nu$ as compared with its value  in the groundstate \eqref{gs_sf}.
Plugging $\rho_0$ into the charge currents in Eq.~\eqref{U(1)-current}, we find that the charge current  $J$ vanishes $J_i=0$. Nevertheless, the dipole currents $\mathcal{D}^{(a)}$ or the many-body current $\mathbf{\Xi}$ in  Eq.~(\ref{CurXi})
which after condensation reads
\begin{equation}
\Xi_{ia}=-2\rho _{0}^2 K_{ia}
\partial _{i}\partial _{a}\theta \,\quad \text{after condensation}, \label{Xi-con}
\end{equation}%
 takes a finite value:
\begin{equation}
\Xi_{s} =\kappa\rho_0^2 \nu\,,
\end{equation}
where $\Xi_{s} =\sqrt{\sum_{i,a=1}^d |\Xi_{ia}|^2}$. 
At $\nu=\sqrt{\frac{g}{3\kappa}}$, $\Xi_{s}$ reaches its maximum $(\Xi_{s})_\mathrm{max}$
\begin{equation}
(\Xi_{s})_\mathrm{max}=\frac{3\sqrt{3\kappa}\mu^2}{16\sqrt{g^3}}\,.
\label{DipileCurrentcrit}
\end{equation}
 {We conclude that, $(\Xi_{s})_\mathrm{max}$ plays the similar role as the critical charge current of a conventional superfluid phase. It means that a fractonic superfluid phase can survive when no induced charge currents appear and $\Xi_{s}$ is smaller than $(\Xi_{s})_\mathrm{max}$ in Eq.~\eqref{DipileCurrentcrit} and that the many-body currents can flow dissipationlessly.} For the diagonal case $K_{ij}=\frac{1}{2}\kappa\delta_{ij}$ under a proper boundary condition, we can obtain the same results as the isotropic case. \textit{Therefore, the  many-body current  $\mathbf{\Xi}$ is the supercurrent of the fractonic superfluid phase.}

%%%%%%%%%%%%%%%
\begin{figure}[tb]
\centering
\includegraphics[scale=0.32]{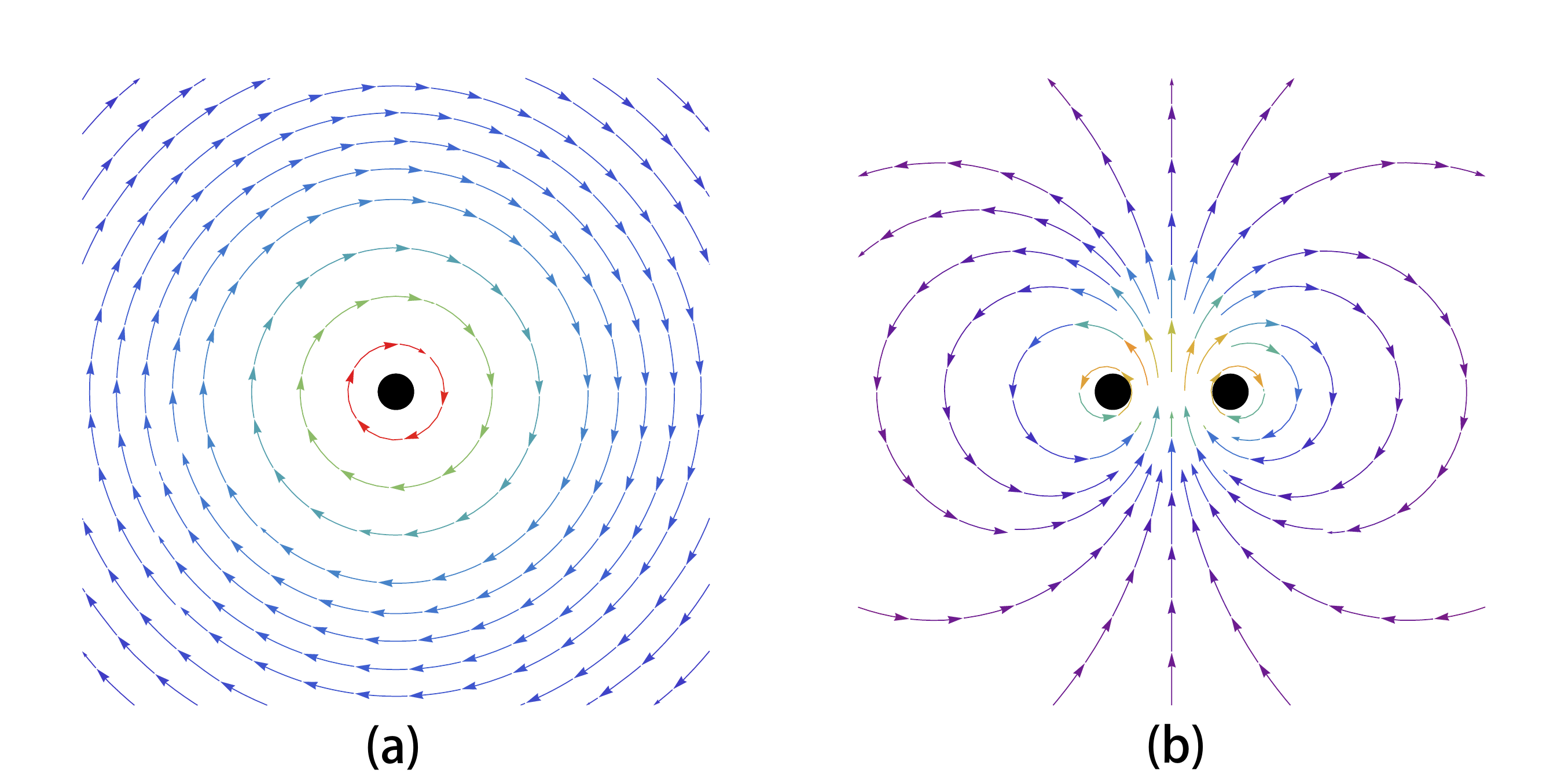}
\caption{Illustration of vector field $\mathbf{U}$ in  Eq.~(\ref{velocity-u})  for (a) a single vortex $\theta=\arctan\frac{x^2}{x^1}$ and (b) a pair of vortex and its anti-vortex,  for simplicity, in 2 spatial dimensions. The color and direction of arrow denote the strength (red$>$blue) and direction of  $\mathbf{U}$. }
\label{Fig:Figure1}
\end{figure}
%%%%%%%%%%%%
%%%%%%%%%%%%
\begin{figure*}[tbp]
\centering
\includegraphics[scale=0.75]{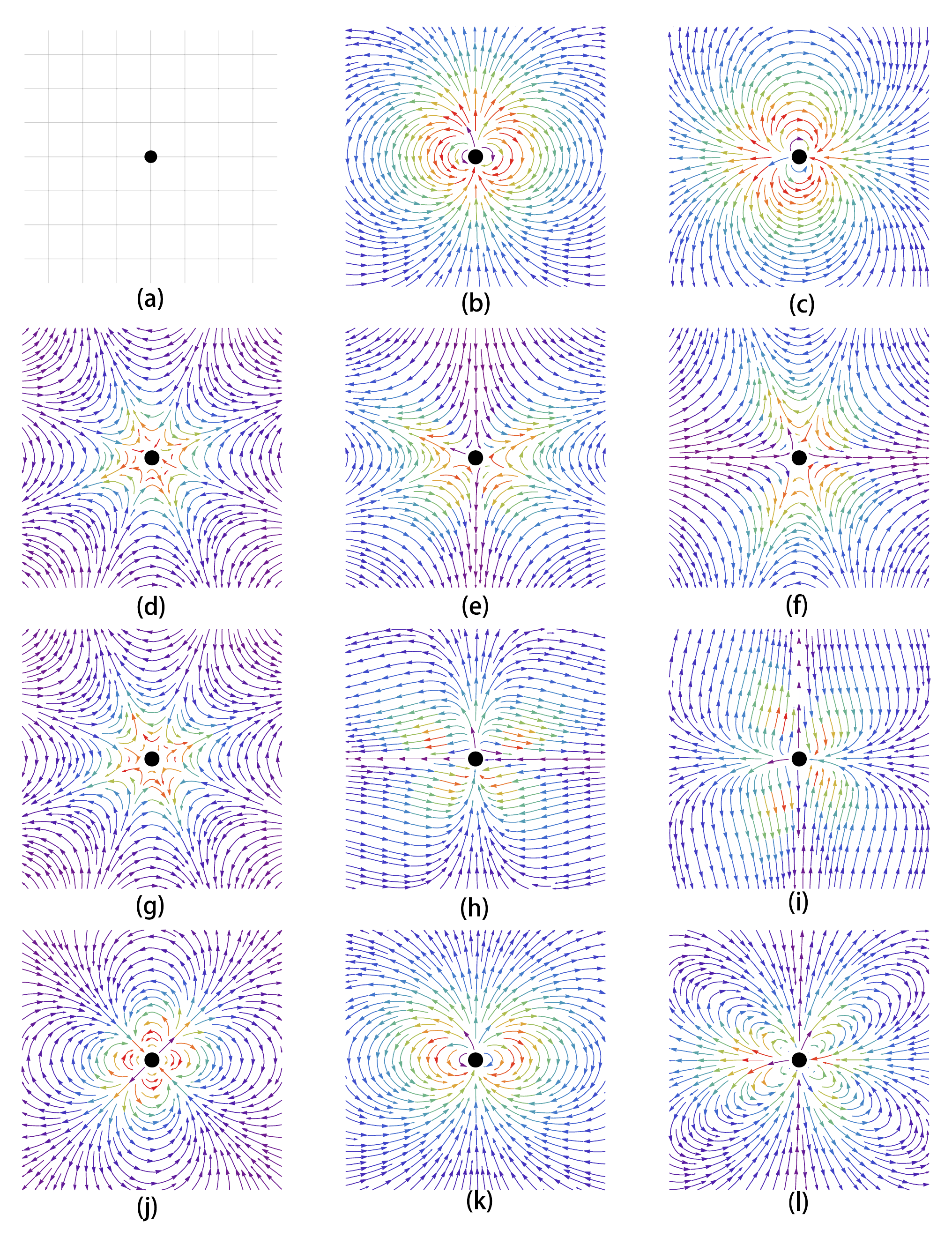}
\caption{Illustration of hydrodynamic fields in fractonic superfluid in the presence of a single vortex, for simplicity, in 2 spatial dimensions.  The direction of an arrow denotes direction of fields  and color characterizes strength. The vortex core in a vortex configuration of $\mathbf{\Xi}$ in Eq.~(\ref{velocity-u}) is marked by a black dot.  The three columns from left to right depict velocity  fields  $\mathbf{v}$ in Eq.~(\ref{velocity-v}), $\mathbf{v}^{(1)}$ and $\mathbf{v}^{(2)}$ in Eq.~ (\ref{velocity-dipole-cond}) respectively.
The four rows from top to bottom correspond to different coupling constants: isotropic  $K_{ij}=\frac{1}{2}\kappa$, 
intermediate $K_{ij}=\left(\begin{array}{cc}
          0.5\kappa  & 0.25\kappa\\
          0.25\kappa & 0.5\kappa
\end{array}\right)$, % $K_{11}=K_{22}=2K_{12}=2K_{21}=\frac{1}{2}\kappa $ 
diagonal  $K_{ij}=\frac{1}{2}\kappa \delta_{ij} $ 
and   $K_{ij}=\left(\begin{matrix}
           1.5\kappa  & \kappa\\
            \kappa & 0.5\kappa
\end{matrix} \right)$
% $K_{11}=2K_{22}=\frac{1}{2}\kappa,K_{12}=K_{21}=2K_{22}=\frac{1}{2}\kappa$
  with a positive constant $\kappa$. The vector fields $\mathbf{\Xi}$ in these cases share the same configuration as in Fig.~\ref{Fig:Figure1}(b).  The charge current for isotropic $K_{ij}$ vanishes in (a) where $\theta$ is an exact solution to Gross-Pitaevskii equation in Eq.~\eqref{Gpequ}.
The vorticities of  charge and dipole velocity fields  are not topological. All the velocity fields are obtained under the same field $\theta=\arctan\frac{x^2}{x^1}$ with winding number $N=1$.}
\label{Fig:vortex}
\end{figure*}
%%%%%%%%%%%%%%%%
\begin{figure*}[tbp]
\centering
\includegraphics[scale=0.75]{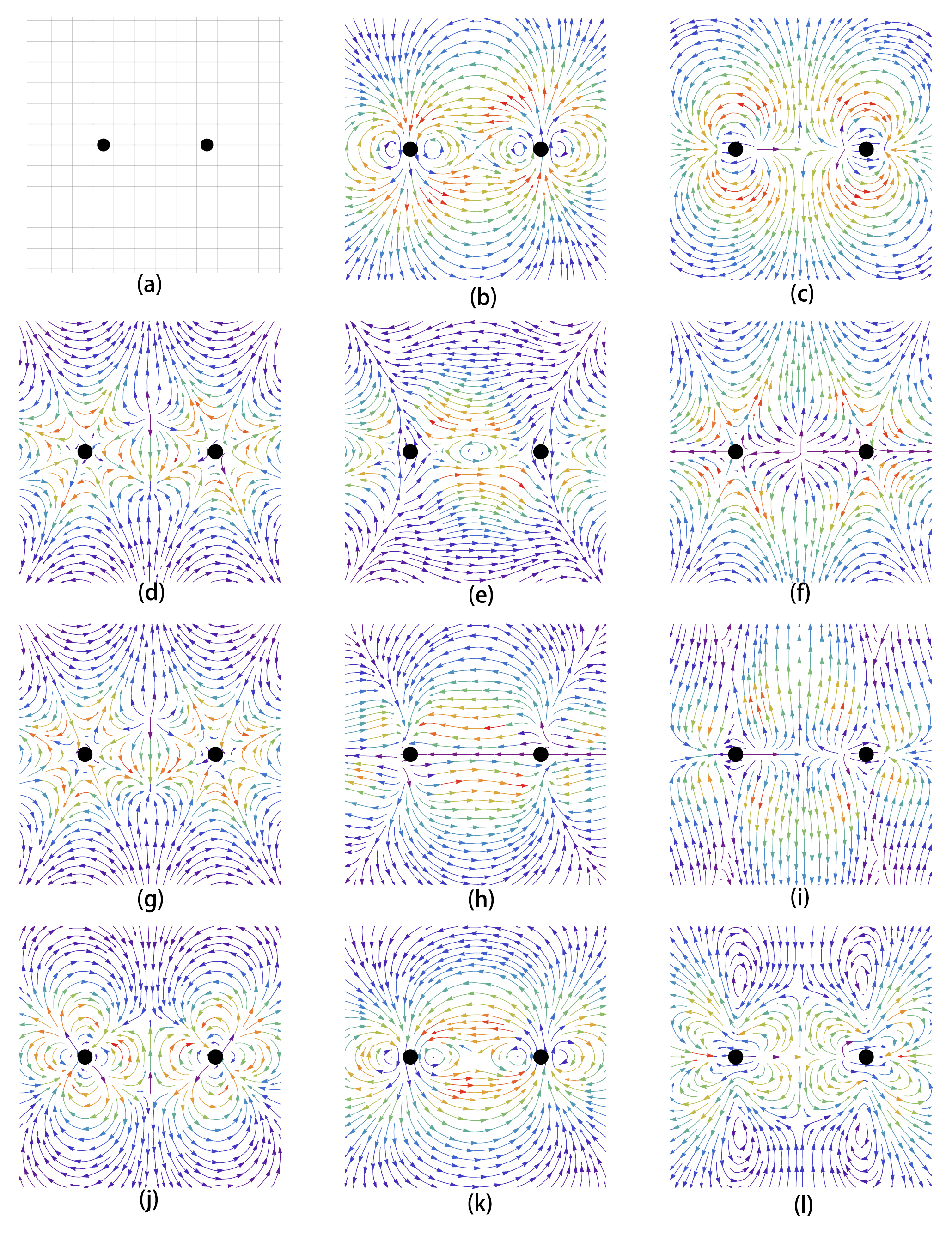}
\caption{Illustration of hydrodynamic fields in fractonic superfluid in the presence of  a pair of vortex-antivortex, for simplicity, in 2 spatial dimensions.  The direction of an arrow denotes direction of fields  and color characterizes strength. The vortex core in a vortex configuration of $\mathbf{\Xi}$ in Eq.~(\ref{velocity-u}) is marked by a black dot.  The three columns from left to right depict velocity  fields  $\mathbf{v}$ in Eq.~(\ref{velocity-v}), $\mathbf{v}^{(1)}$ and $\mathbf{v}^{(2)}$ in Eq.~ (\ref{velocity-dipole-cond}) respectively.
The four rows from top to bottom correspond to different coupling constants as those in Fig.~\ref{Fig:vortex}. 
 The two vortices have  the winding number $1$ (left) and $-1$ (right).  }
\label{Fig:vortex2}
\end{figure*}

%%%
\subsection{Topological vortices}
A conventional $U(1)$ superfluid has vortex excitations and their topological nature can be characterized by vorticity as a close line integral over charge currents that are proportional to $\nabla\theta$.  It is a different story for a fractonic superfluid phase due to its complicated form in Eq.~\eqref{Ham}. Instead, a many-body current $\mathbf{\Xi}$ in Eq.~\eqref{CurXi} that arises from purely two-particle correlated hopping directly gives vortex's topological number.

Given a groundstate wavefunction in Eq.~(\ref{gs_sf}) for a fractonic phase, we are ready
to discuss its topological nature. Our guiding rule is to relate vortex degree of freedom as the singular component of field $\theta$ to a winding number. The velocities in Eqs.~(\ref%
{rank-0-velocity}) and (\ref{rank-1-velocity}) reduce to simpler forms
after condensation with a uniform density distribution $\rho =\rho _{0}$, 
\begin{align}
v_{i} &=-2\rho _{0}\sum_{j}^{d}K_{ij}\partial _{i}\partial _{j}^{2}\theta\,\quad \text{after condensation},
\label{velocity-v} \\
v_{i}^{\left( a\right) }&  =  x^{a}v_{i}+2K_{ia}\rho _{0}\partial _{i}\partial
_{a}\theta \,\quad \text{after condensation}. \label{velocity-dipole-cond}
\end{align}
 Topological number in a
conventional superfluid is embedded in current vorticity. A vorticity can be
expressed as a winding number around the vortex core. Nevertheless, for general $%
K_{ij}$, vorticities of $\mathbf{v}$ and $\mathbf{v}^{\left( a\right) }$ are
no longer topologically invariant. In other words,  $\oint_{C}\mathbf{v}\cdot d%
\mathbf{r}$ and $\oint_{C}\mathbf{v}^{\left( a\right) }\cdot d\mathbf{r}$ 
depend on local geometry of a closed path $C$ since $\mathbf{v}_{i}$ and $\mathbf{v}_{i}^{\left(
a\right) }$ are not \textit{closed} if using the terminology of differential forms. 

As stated in Sec.~\ref{Sec:NoetherThm}, the current $\mathbf{\Xi}$ arises from  a purely many-body effect. 
By contracting one of the two indexes of $\mathbf{\Xi}$ in Eq.~\eqref{Xi-con} after condensation, we can construct a new vector field $\mathbf{U}$
\begin{align}
U_{i}=\sum_{a=1}^{d}K_{ia}^{-1}x^{a}\Xi_{ia} =-\sum_{a=1}^{d}2\rho_0^2 x^{a}\mathbf{\partial }_{i}\partial _{a}\theta \,.
\label{velocity-u}
\end{align}
The prefactor $K_{ia}^{-1}$ is introduced to absorb anisotropy
in $K_{ij}$ and $x^a$ is to decrease degree of derivative. If some $K_{ij}$
vanish, the definition of $\mathbf{\Xi}$ should be understood by taking the
limit $K_{ij}\rightarrow 0$. Thus, Eq.~(\ref{velocity-u}) works for general $%
K_{ij}$. The vorticity $\ell$ associated with $\mathbf{U}$ can be calculated directly, 
\begin{align}
\ell&=\oint\limits_{C}\mathbf{U}\cdot \mathrm{d}\mathbf{r}  \notag
\\
&=-\sum_{a=1}^{d}\oint\limits_{C}\left[ \mathbf{\nabla }\left( x^{a}2\rho
_{0}^2\partial _{a}\theta \right) -\left( \mathbf{\nabla }x^{a}\right) 2\rho
_{0}^2\partial _{a}\theta \right] \cdot \mathrm{d}\mathbf{r}  \notag \\
&=2\rho _{0}^2\oint\limits_{C}\mathbf{\nabla }\theta \cdot d\mathbf{r}=2\rho^2
_{0}2\pi N\,,  \label{vorticity-u}
\end{align}%
where $C$ is a closed loop in $d$ spatial dimensions and $N$ is a summation over winding numbers of vortices surrounded by the loop $C$. Therefore, $\ell$ can be utilized to characterize the
topological nature of vortices. The vanishing of the first term in the
second line of Eq.~(\ref{vorticity-u}) is due to smoothness of $\partial
_{a}\theta $ although $\theta $ is a multivalued function, since the microscopic model in Eq.~\eqref{Ham} appears in a flat-band platform.  
From the  construction of $\Xi$ in Eq.~\eqref{CurXi}, we can conclude that topological properties arise from  a purely many-body effect.

Pictorially, Figs.~\ref{Fig:Figure1}(a) and \ref{Fig:Figure1}(b) show the configurations of  $\mathbf{U}$ of a single vortex and a pair of vortex and anti-vortex  respectively, for simplicity, in two spatial dimensions. In Fig.~\ref{Fig:Figure1}(a) $\mathbf{U}$ circulates around the vortex core marked by a black point. Since we are only concerned about the topological nature, we choose $\theta$ field of a single vortex as $\theta=\arctan x^2/x^1$ that  belongs to the same topological sector with  exact vortex solutions to Gross-Pitaevskii equation in Eq.~\eqref{Gpequ}.  The configurations
of different vector fields in Eqs.~(\ref{velocity-v}), (\ref{velocity-dipole-cond}) and (\ref{velocity-u}) are plotted in Fig.~\ref%
{Fig:vortex}. Obviously, charge currents no longer wind around the vortex core,  which is different from a conventional superfluid phase. Different choices of $K_{ij}$ induce different charge current configurations in Figs.~\ref{Fig:vortex}(d), (g) and (j). Especially, when $\theta$ configuration appears as an exact solution to Gross-Pitaevskii equation in Eq.~\eqref{Gpequ} for isotropic $K_{ij}$ in Fig.~\ref{Fig:vortex}(a), charge currents vanish, which sharpens  the failure of extracting topological number from charge currents. In Fig.~\ref{Fig:vortex2},  the configurations
of different vector fields in Eqs.~(\ref{velocity-v}), (\ref{velocity-dipole-cond}) and (\ref{velocity-u}) of a pair of vortex and anti-vortex (Fig.~\ref{Fig:Figure1}(b)) are depicted.  The current distributions in Fig.~\ref{Fig:vortex2} will be helpful in the analysis of interaction between vortices, which is left to future work.

\bigskip
In conclusion, we identify a fractonic superfluid phase. Different from a conventional
superfluid phase, the topological vortices are characterized by the many-body current $\mathbf{\Xi}$ 
from the pure effect of two particles' correlated hopping and the corresponding supercurrents turn out to be the many-body currents $\mathbf{\Xi}$. Although for simplicity only vortex configurations in two spatial dimensions is discussed, a  vortex in three spatial dimensions that has a line as its core has a similar structure.
In Sec. \ref{Sec:FS_QFT}, we will consider quantum fluctuations against
stability of a fractonic superfluid phase.  
Although quantum fluctuation can destroy superfluidity in a purely two dimensional system, a fractonic superfluid phase may still stay stable in a quasi-two spatial dimensions by the interlayer coupling.  

%%%%%%%%%%%%%%%%%%%%%%%%%%%%%%%%%%%%%%%%

\section{Quantum fluctuations}
\label{Sec:FS_QFT}

Quantum fluctuations can cause instability of a superfluid phase (ie., ODLRO established in classical analysis). In
this section we target on this problem. We first derive an effective theory
for phase fluctuations  based on our microscopic many-fracton model in Eq.~(\ref{Ham}) and then deal with stability of a fractonic
superfluid phase.  

\subsection{Effective theory of the microscopic model: coherence length and effective Lagrangian}

To include quantum fluctuations, without loss of generality, we start with
one classical field configuration $\phi _{0}\left( \mathbf{x},t\right) =%
\sqrt{\rho _{0}}$. Around it, $\phi \left( \mathbf{x},t\right) =\sqrt{\rho
_{0}+\rho \left( \mathbf{x},t\right) }e^{i\theta \left( \mathbf{x},t\right)
} $ where two real fields $\rho \left( \mathbf{x},t\right) $ and $\theta
\left( \mathbf{x},t\right) $ represent density and phase fluctuations
respectively. We have removed the Greek letter $\delta $ in $\delta \rho $
for notation simplicity.  We remark that the field $\theta$ is angular-valued and should be defined $\operatorname{mod} 2\pi$. To the second order, we can derive the effective
Lagrangian corresponding to the microscopic model in Eq.~\eqref{Ham},  
\begin{align}
\mathcal{L} =&-\rho \partial _{t}\theta -\sum_{i,j}^{d}K_{ij}\left[ \rho
_{0}^{2}\left( \partial _{i}\partial _{j}\theta \right) ^{2}+\frac{1}{4}%
\left( \partial _{i}\partial _{j}\rho \right) ^{2}\right]  \notag \\
&-\sum_{i}^{d}G_{i}(\partial _{i}\rho )^{2}-\frac{g}{2}\rho ^{2}.
\label{L2order}
\end{align}%
The density fluctuation field $\rho $ should satisfy a condition as an
auxiliary field $\frac{\delta \mathcal{L}}{\delta \rho }=0$. The solution
takes the form in the momentum space as 
\begin{equation}
\rho \left( \omega ,\mathbf{k}\right) =\frac{-i\omega \theta \left( \omega ,%
\mathbf{k}\right) }{g+\frac{1}{2}\left(
\sum_{i,j}^{d}K_{ij}k_{i}^{2}k_{j}^{2}\right) +2\left(
\sum_{i}^{d}G_{i}k_{i}^{2}\right) }
\end{equation}%
In the long-wave length limit, 
\begin{equation}
g\gg \frac{1}{2}\left( \sum_{i,j}^{d}K_{ij}k_{i}^{2}k_{j}^{2}\right)
+2\left( \sum_{i}^{d}G_{i}k_{i}^{2}\right)  \label{appx_cond}
\end{equation}%
which requires an upper bound for momentum $\left\vert \mathbf{k}\right\vert
\ll 2\pi \xi _{\mathrm{coh}}^{-1}$. $\xi _{\mathrm{coh}}$ is the \textit{coherence length} of the fractonic superfluid and is
determined when right hand and left hand in Eq.~(\ref{appx_cond}) equal. So
we can safely make the approximation: $
\rho (\omega ,\mathbf{k})=-\frac{i\omega \theta (\omega ,\mathbf{k})}{g}\,.
$ Therefore, we obtain
an effective Lagrangian for field $\theta $ 
\begin{equation}
\mathcal{L}=\frac{1}{2g}(\partial _{t}\theta )^{2}-\rho
_{0}^{2}\sum_{i,j}^{d}K_{ij}(\partial _{i}\partial _{j}\theta )^{2}\,.
\label{Goldstone_L}
\end{equation}%
In Eq.~(\ref{Goldstone_L}) since $G_{i}$-term is less relevant, we have
neglected it. The same dispersion relation for
the field $\theta $ can be given through a classical analysis and we will not go into details. There are two issues to be addressed here. First, the  effective theory in Eq.~\eqref{Goldstone_L} is valid only when the length scale is much larger than the coherence length $\xi_{\mathrm{coh}}$. Thus, the prediction power of the effective theory becomes questionable for length scale smaller than or comparable to $\xi_{\mathrm{coh}}$.  Second,  $\theta$ is in fact  an angular-valued field and is defined $\operatorname{mod} 2\pi$. Now we just treat as a real valued field, which is enough for description of the fixed point physics of the fractonic superfluid phase.

Although the broken symmetry $\mathcal{G}$ has $d+1$ generators, we only
have one gapless mode $\theta $ in Eq.~(\ref{Goldstone_L}). Notice that any
vacuum with a broken $U\left( 1\right) $ charge symmetry also is one for
symmetry generated by $Q^{(a)}$ and the main effect of a charge $Q^{\left(
a\right) }$ is to impose a strong constraint on current $J_{i}$ so that the
Goldstone mode has a higher-order dispersion relation.

\subsection{Stability of ODLRO: correlation functions and specific heat}

%%%%%%%%%%%%%%%%%%%%%%%%%%%
\renewcommand\arraystretch{1.7}
\begin{table}[tbp]
\caption{Comparison between correlators in Eq.~(\protect\ref{C(x)}) of conventional superfluid and \textit{isotropic} fractonic superfluid at zero temperature after quantum fluctuations are included. $d$ is spatial dimension. 
The effective theories are given by   Eqs.~(\protect\ref%
{Goldstone-con}) and (\protect\ref{Goldstone-iso}) respectively. The former has $c=\protect\sqrt{\protect\kappa g\protect\rho %
_{0}^{{}}}$ and coherence length $\protect\xi _{\mathrm{coh}}=2\protect\pi\protect\sqrt{%
\protect\kappa /(4\protect\rho _{0}^{{}}g)}$ while the latter with isotropic
coupling constant $K_{ij}=\frac{1}{2}\protect\kappa$ has $c=\protect\sqrt{%
\protect\kappa g\protect\rho _{0}^{2}}$ and coherence length $\protect\xi %
_{\mathrm{coh}}=2\protect\pi \sqrt[4]{\protect\kappa /4g}$. A many-fracton system is fully disordered (marked by $\times$ ) 
  in $d=1$ and algebraically ordered (AO) in $d=2$. It has a stable
ODLRO, i.e., a true superfluid (marked by \checkmark), when $d\geq 3$ at zero temperature. Here $\protect\gamma $ is the
Euler's constant. }
\label{Tab:Correlation}%EndExpansion 
\begin{tabular}{p{0.6cm}p{3cm}p{0.5cm}p{2.8cm}p{0.5cm}}
\specialrule{0.1em}{0pt}{0.6pt}
$d$ & \multicolumn{2}{l}{{Conventional system}} &\multicolumn{2}{l}{{Many-fracton system}} \\
\specialrule{0.05em}{0.5pt}{0.5pt}
$1$ & $\rho _{0}e^{-\frac{\gamma g}{4\pi c}}(\pi r/\xi _{\mathrm{coh}})^{-\frac{g}{2\pi
c}}$ & AO & $\rho _{0}e^{-\frac{g}{2c}( \pi r-\xi _{\mathrm{coh}}/\pi ^{\frac{3}{2}%
}) }$ & $\times$ \\ 
$2$ & $\rho _{0}e^{-\frac{g}{2\pi c}\xi _{\mathrm{coh}}^{-1}}$ & \checkmark & $\rho _{0}e^{-%
\frac{\gamma g}{4\pi c}}(r/\xi _{\mathrm{coh}})^{-\frac{g}{2\pi c}}$ & AO \\ 
$\geq 3$ & $\rho _{0}e^{-g\frac{\pi ^{\frac{d-3}{2}}}{2\left( d-1\right) c}%
\xi _{\mathrm{coh}}^{1-d}}$ & \checkmark & $\rho _{0}e^{-\frac{g}{c}\frac{\pi ^{\frac{d}{2}%
-2}}{2(d-2)}\xi _{\mathrm{coh}}^{2-d}}$ & \checkmark \\
\specialrule{0.1em}{1pt}{0pt}
\end{tabular}%
\end{table}
%
%%%%%%%%%%%%%%%%%%%%%%

We are ready to include the effect of quantum fluctuations on
correlation $C\!\left( \mathbf{x}\right) $ in Eq.~(\ref{correlation_cl}).
The correlator $C\!\left( \mathbf{x}\right) $ is related to an equal-time
Green function of the Golstone mode $\theta \!\left( \mathbf{x},0\right) $ 
\begin{equation}
C\left( \mathbf{x}\right) =\rho _{0}e^{-\frac{1}{2}\left\langle \left[ \theta \left( \mathbf{x}%
,0\right) -\theta \left( 0,0\right) \right] ^{2}\right\rangle } \,, \label{C(x)}
\end{equation}%
where the equal-time Green function can be calculated in the momentum space: 
\begin{align}
&\left\langle \left[ \theta \left( \mathbf{x},0\right) -\theta \left(
0,0\right) \right] ^{2}\right\rangle  \notag \\
=&\int \frac{d^{d}\mathbf{k}d\omega }{\left( 2\pi \right) ^{d+1}}\left( 2-2e^{-i%
\mathbf{k}\cdot \mathbf{x}}\right) \left\langle \theta \!\left( \omega ,%
\mathbf{k}\right) \theta \!\left( -\omega ,-\mathbf{k}\right) \right\rangle\,.
\label{theta_cor}
\end{align}%
Eq.~(\ref{theta_cor}) is hard to deal with exactly for a general $K_{ij}$.
In the following, we consider two specified cases.

\textbf{Isotropic case $K_{ij}=\frac{1}{2}\kappa $ for any $i,j$}.---
We introduce a factor $\frac{1}{2}$ in $K_{ij}$ to simplify our expressions.
In this case, the Goldstone mode $\theta $ has a quadratic dispersion
relation 
\begin{equation}
\omega =\sqrt{\kappa g\rho _{0}^{2}}\left\vert \mathbf{k}\right\vert
^{2}\equiv c\left\vert \mathbf{k}\right\vert ^{2}  \label{dispersion-isotro}
\end{equation}%
and it recovers a rotational symmetry. The coherence length $\xi _{\mathrm{coh}}$ is
determined by equation $g=\frac{1}{4}\kappa \left( \frac{2\pi }{\xi _{\mathrm{coh}}}%
\right) ^{4}$ where $G_{i}$-term is less relevant for the upper bound of
momentum. So we have%
\begin{equation}
\xi _{\mathrm{coh}}=2\pi \left( \frac{\kappa }{4g}\right) ^{\frac{1}{4}}\,.
\label{cohlengtiso}
\end{equation}%
The Lagrangian 
\begin{equation}
\mathcal{L}=\frac{1}{2g}(\partial _{t}\theta )^{2}-\frac{1}{2g}c^{2}\left(
\nabla ^{2}\theta \right) ^{2}  \label{Goldstone-iso}
\end{equation}%
possesses a Lifshitz spacetime symmetry and is related to nonrelativistic
gravity\cite{Hoava2009, Hoava2009a, Xu2010}.

 The asymptotic behavior of $%
C\left( \mathbf{x}\right) $ has been listed in Table \ref{Tab:Correlation}.
We can find that
only when our space dimension $d>2$ does a superfluid survive quantum
fluctuations. The correlator $C\left( \mathbf{x}\right) $ approaches zero in
dimension $d=1$ and $2$ in large distance. We point out that $C\left( 
\mathbf{x}\right) $ decays in a power-law pattern in $d=2$, which is similar
to a conventional superfluid in $d=1$. Another
aspect of the higher-order dispersion in Eq.~(\ref{dispersion-isotro}) is
specific heat capacity:%
%\begin{align}
%c_{V} &=\frac{1}{V}\frac{dE}{dT}  \notag \\
%&=(T/c)^{\frac{d}{2}}\frac{\Omega _{d-1}}{2(2\pi )^{d}}\int_{0}^{\infty }%
%\frac{x^{\frac{d+2}{2}}e^{x}}{\left( e^{x}-1\right) ^{2}}dx\,,
%\end{align}%
\begin{align}
c_{v}^{} =(T/c)^{\frac{d}{2}}\frac{\Omega _{d-1}}{2(2\pi )^{d}}\Gamma(\frac{d}{2}+2)\zeta(\frac{d+2}{2})~,
\label{specific-heat-cap}
\end{align}%
where  $\Omega _{d-1}$ is surface area of unit $\left( d-1\right) $%
-sphere and $\zeta(s)$ is the Riemann zeta function. From Eq.~(\ref{specific-heat-cap}), $c_{v}$ is proportional to $T^{%
\frac{d}{2}}$ in $d$ space dimensions. When a spatial dimension is lower
than $3$, the specific heat capacity is physically meaningless. The result
in Eq.~(\ref{specific-heat-cap}) for $d=3$ is valid under our assumption of
existence of Goldstone mode at finite temperature. On the other hand, for a
conventional superfluid with Lagrangian $\mathcal{L}=i\phi ^{\ast }\partial
_{t}\phi -\frac{1}{2}\kappa \left\vert \nabla \phi \right\vert ^{2}-V\left(
\phi \right) $ with $V\left( \phi \right) $ in Eq.~(\ref{g-potential}), the
effective theory for the Goldstone mode is 
\begin{equation}
\mathcal{L}=\frac{1}{2g}(\partial _{t}\theta )^{2}-\frac{1}{2g}c^{2}\left(
\nabla \theta \right) ^{2} \,, \label{Goldstone-con}
\end{equation}%
where Goldstone mode has a linear dispersion relation $\omega =\sqrt{kg\rho
_{0}}\left\vert \mathbf{k}\right\vert \equiv c\left\vert \mathbf{k}%
\right\vert $ and quantum fluctuation will kill a superfluid phase in one
spatial dimension at zero temperature. Table \ref{Tab:Correlation} makes a
comparison between a conventional and fractonic superfluid phase. Interestingly, 
Ref.~[\onlinecite{MaHigherRankDQC}] discussed a similar effective Lagrangian from a different higher-derivative model where   excitons \cite{FisherEBL} form a condensate. In our context, the effective Lagrangian in Eq.~(\ref{Goldstone-iso}) is a description of fractonic superfluids in the isotropic case, whose microscopic origin is given by Eq.~(\ref{Ham}) and the  order parameter is given by $\langle\hat{\Phi}\rangle$ in Table~\ref{Tab_key_result}.

\textbf{Diagonal case $K_{ij}=\frac{1}{2}\kappa \delta _{ij}$}.---
Now the Goldstone mode $\theta $ has a dispersion spectrum 
\begin{equation}
\omega =\sqrt{\kappa g\rho _{0}^{2}}\sqrt{\sum_{i}^{d}k_{i}^{4}} \equiv c\sqrt{ \sum_{i}^{d}k_{i}^{4}}\,.
\end{equation}%
It does not have a rotational symmetry. We can arrive at the same conclusion
as the isotropic case.

The above analysis just demonstrates that a superfluid phase or ODLRO cannot
survive against quantum fluctuation when the spatial dimension is lower than 
three at zero temperature.
 In particular, in $d=2$ quantum fluctuations only allow an algebraic order. However, similar to superconductivity in 2d materials, interlayer couplings or a quasi-2d system can stabilize the fractonic superfluid against quantum fluctuations.  
 Thermal effect may destroy a fractonic superfluid
phase and the related results will be present in future work. In Appendix.~\ref%
{Sec:General-model}, we aim to discuss   general many-fracton models.

\section{Conclusion}

\label{Sec:Conclusion}
In this paper, we have studied a many-fracton model in the microscopic Hamiltonian (\ref{Ham}) that
lacks of mobility of a single particle. The model in Eq.~(\ref{Ham})
conserves both charge and total dipole moments. We have derived nontrivial
Euler-Lagrange equation and the Noether currents. By taking a Mexican-hat
potential, we deduce a time-dependent Gross-Pitaevskii-type equations. We finally end up with fractonic superfluidity from both classical and quantum levels of length scale much larger than coherence length $\xi_{\mathrm{coh}}$, including supercurrent, topological vortices, ODLRO stability against gapless Goldstone modes and specific heat in low temperatures. The Hamiltonian in Eq.~\eqref{Ham} is expected to be realized in cold atomic gas subjected to an optical lattice \cite{ExperimentFisher2005,2008PhRvA77b3603P}, especially when a trap is considered, and opens a new horizon to search for   exotic phases of matter.
There are many interesting directions for future investigation. 
For example, we can discuss a \textit{fractonic superconducting}
phase by allowing  a fracton field $\hat{\Phi}$  in Eq.~\eqref{Ham}  to satisfy  anti-commutation relations with possible pairing field $\Delta_{ij}\sim \hat{\Phi}\partial_i\partial_j\hat{\Phi}-\partial_i\hat{\Phi}\partial_j\hat{\Phi}$.
And then, the BEC-BCS crossover of a fracton system may show interesting
physical consequence. Despite of highly non-linearity, numerically solving the Gross-Pitaevskii equations will be very attractive.   
One can consider fractonic versions of other types of ordered phases, such as nematic and stripe orders and discuss their competitions.
 One may also consider a \emph{symmetric} phase formed by condensing unconventional vortices in the fractonic superfluid. By decorating on-site symmetry charge on vortices, one may construct SPTs with both higher rank symmetry and usual on-site symmetry, following the similar methods in usual   SPT constructions \cite{Chen:2014aa,PhysRevB.93.115136,YeGu2015,bti2,yp18prl}.

\section*{Acknowledgements}
We thank Meng Cheng and Juven Wang for useful discussions. The work was supported by the SYSU startup grant and NSFC no. 11847608.

\appendix
 
\section{General many-fracton models}

\label{Sec:General-model} We can generalize the many-fracton model in Eq.~(\ref{Ham}) into a large class. We begin with a Hamiltonian $H=\int \mathrm{d}^{d}\mathbf{x}\mathcal{H}_N$
where the Hamiltonian density $\mathcal{H}_N$ reads 
\begin{widetext}
\begin{equation}
\mathcal{H}_N=\sum_{i_{1}i_{2}\cdots i_{N+1}}^dK_{i_{1}i_{2}\cdots i_{N+1}}\left(\hat{\Phi} ^{\dag }\right)
^{N+1}\left( \nabla _{i_{1}i_{2}\cdots
i_{N+1}}^{{}}\log\hat{\Phi} ^{\dag }\right)\hat{\Phi} ^{N+1}\left( \nabla
_{i_{1}i_{2}\cdots i_{N+1}}^{{}}\log\hat{\Phi} \right) +V\!\left(\hat{\Phi} ^{\dag
},\hat{\Phi} \right)   \,,\label{General-Ham}
\end{equation}%
\end{widetext}
where $\nabla _{i_{1}i_{2}\cdots i_{n}}^{{}}=\partial _{i_{1}}\partial
_{i_{2}}\cdots \partial _{i_{n}}$ and the summation for each index is over
all spatial dimensions. The coupling constant $K_{i_{1}i_{2}\cdots i_{N+1}}$
can be anisotropic and it is fully symmetric with its indexes. When $N=0$,
Eq.~(\ref{General-Ham}) reduces to a Gaussian theory and when $N=1$, it
reduces to the many-fracton model in Eq.~(\ref{Ham}) except $G_{i}$-term.
Here some models are listed 
\begin{widetext}
\begin{align}
\mathcal{H}_{0} =&\sum_{i}^{d}K_{i}\partial _{i}\hat{\Phi} ^{\dag }\partial _{i}\hat{\Phi}
+V\!\left(\hat{\Phi} ^{\dag },\hat{\Phi} \right)  \\
\mathcal{H}_{2} =&\sum_{i,j,k}^{d}K_{ijk}\left[ 2\partial _{i}\hat{\Phi} ^{\dag }\partial
_{j}\hat{\Phi} ^{\dag }\partial _{k}\hat{\Phi} ^{\dag }-3\hat{\Phi} ^{\dag }\partial _{i}\hat{\Phi}
^{\dag }\partial _{j}\partial _{k}\hat{\Phi} ^{\dag }+\left(\hat{\Phi} ^{\dag }\right)
^{2}\partial _{i}\partial _{j}\partial _{k}\hat{\Phi} ^{\dag }\right] \notag \\
&\cdot  \left[ 2\partial _{i}\hat{\Phi} \partial _{j}\hat{\Phi} \partial _{k}\hat{\Phi} -3\hat{\Phi}
\partial _{i}\hat{\Phi} \partial _{j}\partial _{k}\hat{\Phi}   +\hat{\Phi} 
^{2}\partial _{i}\partial _{j}\partial _{k}\hat{\Phi} \right]  +V\!\left( \hat{\Phi} ^{\dag
},\hat{\Phi} \right)  \\
\mathcal{H}_{3} =&\sum_{i,j,k,l}^{d}K_{ijkl}\left[ 6\partial _{i}\hat{\Phi} ^{\dag
}\partial _{j}\hat{\Phi} ^{\dag }\partial _{k}\hat{\Phi} ^{\dag }\partial _{l}\hat{\Phi}
^{\dag }-12\hat{\Phi} ^{\dag }\partial _{i}\hat{\Phi} ^{\dag }\partial _{j}\hat{\Phi} ^{\dag
}\partial _{k}\partial _{l}\hat{\Phi} ^{\dag }+4\left( \hat{\Phi} ^{\dag }\right)
^{2}\partial _{i}\hat{\Phi} ^{\dag }\partial _{j}\partial _{k}\partial _{l}\hat{\Phi}
^{\dag }\right.   \notag \\
&\left. +\left( \hat{\Phi} ^{\dag }\right) ^{2}\left( 3\partial _{i}\partial
_{j}\hat{\Phi} ^{\dag }\partial _{k}\partial _{l}\hat{\Phi} ^{\dag }-\hat{\Phi} ^{\dag
}\partial _{i}\partial _{j}\partial _{k}\partial _{l}\hat{\Phi} ^{\dag }\right) 
\right] \left[ 6\partial _{i}\hat{\Phi} \partial _{j}\hat{\Phi} \partial _{k}\hat{\Phi}
\partial _{l}\hat{\Phi} -12\hat{\Phi} \partial _{i}\hat{\Phi} \partial _{j}\hat{\Phi} \partial
_{k}\partial _{l}\hat{\Phi} \right.   \notag \\
&\left. +4\hat{\Phi} ^{2}\partial _{i}\hat{\Phi} \partial _{j}\partial _{k}\partial
_{l}\hat{\Phi} +\hat{\Phi} ^{2}\left(3 \partial _{i}\partial _{j}\hat{\Phi} \partial
_{k}\partial _{l}\hat{\Phi} -\hat{\Phi} \partial _{i}\partial _{j}\partial _{k}\partial
_{l}\hat{\Phi} \right) \right] +V\!( \hat{\Phi} ^{\dag },\hat{\Phi} ) \,.
\end{align}
\end{widetext}
Under the standard coherent-state path integral, we can write down the
Lagrangian $\mathcal{L}=i\phi ^{\ast }\partial _{t}\phi -\mathcal{H}(\phi
^{\ast },\phi )$. Although $\log \phi $ is a multivalued function and has
singularity, the kinetic term turns out to be well-defined.
The system in Eq.~(\ref{General-Ham}) is invariant under a transformation%
\begin{equation}
\phi \rightarrow \exp \left( i\delta \theta \right) \phi\,,
\end{equation}%
where $\delta \theta $ is polynomials of degree $N$ of local coordinates 
\begin{align}
\!\!\!\delta \theta\! =\!\sum_{i_{1}\cdots i_{N}}\!\mathcal{D} _{i_{1}i_{2}\cdots
i_{N}}x^{i_{1}}x^{i_{2}}\cdots x^{i_{N}}+\!\cdots \!+\sum_{i}\mathcal{D}
_{i}x^{i}+\mathcal{D}  \,,\!\!\label{trans-rank-n}
\end{align}%
where $\mathcal{D} _{i_{1}\cdots i_{l}}$ is a symmetric real tensor of rank-$l$
with respect to spatial indexes.  And the related conserved charges have the
form as 
\begin{equation}
Q^{\left( C\left( x^{a}\right) \right) }=\int \mathrm{d}^{d}\mathbf{x}\rho C\left(
x^{a}\right) \,, \label{charge-gener}
\end{equation}%
where $C\left( x^{a}\right) $ is as a homogeneous polynomials with degree-$p$
and $p\leq N$. We dub a symmetry generated by charges in Eq.~(\ref%
{charge-gener}) as a \emph{rank-}$N$\emph{\ symmetry}\cite%
{PhysRevX.9.031035, Seiberg2019arXiv1909}. In this sense, a global $U\left(
1\right) $ is a rank-$0$ symmetry and $\mathcal{G}$ for Hamiltonian in Eq.~(%
\ref{Ham}) is a rank-$1$ symmetry.

Now we focus on an isotropic coupling constant $\mathcal{D}_{i_{1}i_{2}\cdots i_{N+1}}=%
\frac{1}{2}\kappa $. If we take a Mexican-hat potential chemical potential $%
\mu >0$, we have degenerate vacuum with finite uniform density
distribution $\rho =\rho _{0}$. Through the same processes, we can derive an
effective theory for the quantum fluctuation field $\theta $ after
condensation, %
\begin{align}
\mathcal{L}=\frac{1}{2g}\left( \partial _{t}\theta \right) ^{2}-\frac{1}{2g}%
c^{2}\left( \nabla ^{N+1}\theta \right) ^{2}.  \label{rank-N+1-theta}
\end{align}
The effective theory describes Goldstone mode $\theta $ with a dispersion
relation $\omega =\sqrt{\kappa g\rho _{0}^{N+1}}\left\vert \mathbf{k}%
\right\vert ^{N+1}\equiv c\left\vert \mathbf{k}\right\vert ^{N+1}$. The
calculation on the correlator $C\left( \mathbf{x}\right) $
% can be found in
 shows that $C\left( \mathbf{x}%
\right) $ decays to zero when spatial dimension is lower than $d<N+2$ at
zero temperature. In particular, $C\left( \mathbf{x}\right) $ decays in a
power-law pattern at zero temperature at spatial dimension $d=N+1$.

%\bibliographystyle{apsrev4-1}
%\bibliography{top}

\begin{thebibliography}{72}%
\makeatletter
\providecommand \@ifxundefined [1]{%
 \@ifx{#1\undefined}
}%
\providecommand \@ifnum [1]{%
 \ifnum #1\expandafter \@firstoftwo
 \else \expandafter \@secondoftwo
 \fi
}%
\providecommand \@ifx [1]{%
 \ifx #1\expandafter \@firstoftwo
 \else \expandafter \@secondoftwo
 \fi
}%
\providecommand \natexlab [1]{#1}%
\providecommand \enquote  [1]{``#1''}%
\providecommand \bibnamefont  [1]{#1}%
\providecommand \bibfnamefont [1]{#1}%
\providecommand \citenamefont [1]{#1}%
\providecommand \href@noop [0]{\@secondoftwo}%
\providecommand \href [0]{\begingroup \@sanitize@url \@href}%
\providecommand \@href[1]{\@@startlink{#1}\@@href}%
\providecommand \@@href[1]{\endgroup#1\@@endlink}%
\providecommand \@sanitize@url [0]{\catcode `\\12\catcode `\$12\catcode
  `\&12\catcode `\#12\catcode `\^12\catcode `\_12\catcode `\%12\relax}%
\providecommand \@@startlink[1]{}%
\providecommand \@@endlink[0]{}%
\providecommand \url  [0]{\begingroup\@sanitize@url \@url }%
\providecommand \@url [1]{\endgroup\@href {#1}{\urlprefix }}%
\providecommand \urlprefix  [0]{URL }%
\providecommand \Eprint [0]{\href }%
\providecommand \doibase [0]{http://dx.doi.org/}%
\providecommand \selectlanguage [0]{\@gobble}%
\providecommand \bibinfo  [0]{\@secondoftwo}%
\providecommand \bibfield  [0]{\@secondoftwo}%
\providecommand \translation [1]{[#1]}%
\providecommand \BibitemOpen [0]{}%
\providecommand \bibitemStop [0]{}%
\providecommand \bibitemNoStop [0]{.\EOS\space}%
\providecommand \EOS [0]{\spacefactor3000\relax}%
\providecommand \BibitemShut  [1]{\csname bibitem#1\endcsname}%
\let\auto@bib@innerbib\@empty
%</preamble>
\bibitem [{\citenamefont {Pethick}\ and\ \citenamefont
  {Smith}(2008)}]{becbook}%
  \BibitemOpen
  \bibfield  {author} {\bibinfo {author} {\bibfnamefont {C.~J.}\ \bibnamefont
  {Pethick}}\ and\ \bibinfo {author} {\bibfnamefont {H.}~\bibnamefont
  {Smith}},\ }\href@noop {} {\emph {\bibinfo {title} {Bose-Einstein
  Condensation in Dilute Gases}}}\ (\bibinfo  {publisher} {Cambridge University
  Press},\ \bibinfo {year} {2008})\BibitemShut {NoStop}%
\bibitem [{\citenamefont {Chaikin}\ and\ \citenamefont
  {Lubensky}(2000)}]{chaikin2000principles}%
   \BibitemOpen
  \bibfield  {author} {\bibinfo {author} {\bibfnamefont {Paul~M}\ \bibnamefont
  {Chaikin}}\ and\ \bibinfo {author} {\bibfnamefont {Tom~C}\ \bibnamefont
  {Lubensky}},\ }\href@noop {} {\emph {\bibinfo {title} {Principles of
  condensed matter physics}}},\ Vol.~\bibinfo {volume} {1}\ (\bibinfo
  {publisher} {Cambridge university press Cambridge},\ \bibinfo {year}
  {2000})\BibitemShut {NoStop}%
\bibitem [{\citenamefont {Yang}(1962)}]{YangODLRO}%
  \BibitemOpen
  \bibfield  {author} {\bibinfo {author} {\bibfnamefont {C.~N.}\ \bibnamefont
  {Yang}},\ }\bibfield  {title} {\enquote {\bibinfo {title} {Concept of
  off-diagonal long-range order and the quantum phases of liquid he and of
  superconductors},}\ }\href {\doibase 10.1103/RevModPhys.34.694} {\bibfield
  {journal} {\bibinfo  {journal} {Rev. Mod. Phys.}\ }\textbf {\bibinfo {volume}
  {34}},\ \bibinfo {pages} {694--704} (\bibinfo {year} {1962})}\BibitemShut
  {NoStop}%
\bibitem [{\citenamefont {Leggett}(2001)}]{RevModPhys73307}%
  \BibitemOpen
  \bibfield  {author} {\bibinfo {author} {\bibfnamefont {Anthony~J.}\
  \bibnamefont {Leggett}},\ }\bibfield  {title} {\enquote {\bibinfo {title}
  {Bose-einstein condensation in the alkali gases: Some fundamental
  concepts},}\ }\href {\doibase 10.1103/RevModPhys.73.307} {\bibfield
  {journal} {\bibinfo  {journal} {Rev. Mod. Phys.}\ }\textbf {\bibinfo {volume}
  {73}},\ \bibinfo {pages} {307--356} (\bibinfo {year} {2001})}\BibitemShut
  {NoStop}%
\bibitem [{\citenamefont {Bloch}\ \emph {et~al.}(2008)\citenamefont {Bloch},
  \citenamefont {Dalibard},\ and\ \citenamefont {Zwerger}}]{RevModPhys80885}%
  \BibitemOpen
  \bibfield  {author} {\bibinfo {author} {\bibfnamefont {Immanuel}\
  \bibnamefont {Bloch}}, \bibinfo {author} {\bibfnamefont {Jean}\ \bibnamefont
  {Dalibard}}, \ and\ \bibinfo {author} {\bibfnamefont {Wilhelm}\ \bibnamefont
  {Zwerger}},\ }\bibfield  {title} {\enquote {\bibinfo {title} {Many-body
  physics with ultracold gases},}\ }\href {\doibase 10.1103/RevModPhys.80.885}
  {\bibfield  {journal} {\bibinfo  {journal} {Rev. Mod. Phys.}\ }\textbf
  {\bibinfo {volume} {80}},\ \bibinfo {pages} {885--964} (\bibinfo {year}
  {2008})}\BibitemShut {NoStop}%
\bibitem [{\citenamefont {Giorgini}\ \emph {et~al.}(2008)\citenamefont
  {Giorgini}, \citenamefont {Pitaevskii},\ and\ \citenamefont
  {Stringari}}]{RevModPFermigas}%
  \BibitemOpen
  \bibfield  {author} {\bibinfo {author} {\bibfnamefont {Stefano}\ \bibnamefont
  {Giorgini}}, \bibinfo {author} {\bibfnamefont {Lev~P.}\ \bibnamefont
  {Pitaevskii}}, \ and\ \bibinfo {author} {\bibfnamefont {Sandro}\ \bibnamefont
  {Stringari}},\ }\bibfield  {title} {\enquote {\bibinfo {title} {Theory of
  ultracold atomic fermi gases},}\ }\href {\doibase 10.1103/RevModPhys.80.1215}
  {\bibfield  {journal} {\bibinfo  {journal} {Rev. Mod. Phys.}\ }\textbf
  {\bibinfo {volume} {80}},\ \bibinfo {pages} {1215--1274} (\bibinfo {year}
  {2008})}\BibitemShut {NoStop}%
\bibitem [{\citenamefont {{Lewenstein}}\ \emph {et~al.}(2007)\citenamefont
  {{Lewenstein}}, \citenamefont {{Sanpera}}, \citenamefont {{Ahufinger}},
  \citenamefont {{Damski}}, \citenamefont {{Sen}},\ and\ \citenamefont
  {{Sen}}}]{2007AdPhy56243L}%
  \BibitemOpen
  \bibfield  {author} {\bibinfo {author} {\bibfnamefont {Maciej}\ \bibnamefont
  {{Lewenstein}}}, \bibinfo {author} {\bibfnamefont {Anna}\ \bibnamefont
  {{Sanpera}}}, \bibinfo {author} {\bibfnamefont {Veronica}\ \bibnamefont
  {{Ahufinger}}}, \bibinfo {author} {\bibfnamefont {Bogdan}\ \bibnamefont
  {{Damski}}}, \bibinfo {author} {\bibfnamefont {Aditi}\ \bibnamefont {{Sen}}},
  \ and\ \bibinfo {author} {\bibfnamefont {Ujjwal}\ \bibnamefont {{Sen}}},\
  }\bibfield  {title} {\enquote {\bibinfo {title} {{Ultracold atomic gases in
  optical lattices: mimicking condensed matter physics and beyond}},}\ }\href
  {\doibase 10.1080/00018730701223200} {\bibfield  {journal} {\bibinfo
  {journal} {Advances in Physics}\ }\textbf {\bibinfo {volume} {56}},\ \bibinfo
  {pages} {243--379} (\bibinfo {year} {2007})},\ \Eprint
  {http://arxiv.org/abs/cond-mat/0606771} {arXiv:cond-mat/0606771
  [cond-mat.other]} \BibitemShut {NoStop}%
\bibitem [{\citenamefont {{Carr}}\ \emph {et~al.}(2009)\citenamefont {{Carr}},
  \citenamefont {{DeMille}}, \citenamefont {{Krems}},\ and\ \citenamefont
  {{Ye}}}]{2009NJPh11e5049C}%
  \BibitemOpen
  \bibfield  {author} {\bibinfo {author} {\bibfnamefont {Lincoln~D.}\
  \bibnamefont {{Carr}}}, \bibinfo {author} {\bibfnamefont {David}\
  \bibnamefont {{DeMille}}}, \bibinfo {author} {\bibfnamefont {Roman~V.}\
  \bibnamefont {{Krems}}}, \ and\ \bibinfo {author} {\bibfnamefont {Jun}\
  \bibnamefont {{Ye}}},\ }\bibfield  {title} {\enquote {\bibinfo {title} {{Cold
  and ultracold molecules: science, technology and applications}},}\ }\href
  {\doibase 10.1088/1367-2630/11/5/055049} {\bibfield  {journal} {\bibinfo
  {journal} {New Journal of Physics}\ }\textbf {\bibinfo {volume} {11}},\
  \bibinfo {eid} {055049} (\bibinfo {year} {2009})},\ \Eprint
  {http://arxiv.org/abs/0904.3175} {arXiv:0904.3175 [quant-ph]} \BibitemShut
  {NoStop}%
\bibitem [{\citenamefont {Dalibard}\ \emph {et~al.}(2011)\citenamefont
  {Dalibard}, \citenamefont {Gerbier}, \citenamefont
  {Juzeli\ifmmode~\bar{u}\else \={u}\fi{}nas},\ and\ \citenamefont
  {\"Ohberg}}]{RevModPhys831523}%
  \BibitemOpen
  \bibfield  {author} {\bibinfo {author} {\bibfnamefont {Jean}\ \bibnamefont
  {Dalibard}}, \bibinfo {author} {\bibfnamefont {Fabrice}\ \bibnamefont
  {Gerbier}}, \bibinfo {author} {\bibfnamefont {Gediminas}\ \bibnamefont
  {Juzeli\ifmmode~\bar{u}\else \={u}\fi{}nas}}, \ and\ \bibinfo {author}
  {\bibfnamefont {Patrik}\ \bibnamefont {\"Ohberg}},\ }\bibfield  {title}
  {\enquote {\bibinfo {title} {Colloquium: Artificial gauge potentials for
  neutral atoms},}\ }\href {\doibase 10.1103/RevModPhys.83.1523} {\bibfield
  {journal} {\bibinfo  {journal} {Rev. Mod. Phys.}\ }\textbf {\bibinfo {volume}
  {83}},\ \bibinfo {pages} {1523--1543} (\bibinfo {year} {2011})}\BibitemShut
  {NoStop}%
\bibitem [{\citenamefont {{Georgescu}}\ \emph {et~al.}(2014)\citenamefont
  {{Georgescu}}, \citenamefont {{Ashhab}},\ and\ \citenamefont
  {{Nori}}}]{2014RvMP86153G}%
  \BibitemOpen
  \bibfield  {author} {\bibinfo {author} {\bibfnamefont {I.~M.}\ \bibnamefont
  {{Georgescu}}}, \bibinfo {author} {\bibfnamefont {S.}~\bibnamefont
  {{Ashhab}}}, \ and\ \bibinfo {author} {\bibfnamefont {Franco}\ \bibnamefont
  {{Nori}}},\ }\bibfield  {title} {\enquote {\bibinfo {title} {{Quantum
  simulation}},}\ }\href {\doibase 10.1103/RevModPhys.86.153} {\bibfield
  {journal} {\bibinfo  {journal} {Reviews of Modern Physics}\ }\textbf
  {\bibinfo {volume} {86}},\ \bibinfo {pages} {153--185} (\bibinfo {year}
  {2014})},\ \Eprint {http://arxiv.org/abs/1308.6253} {arXiv:1308.6253
  [quant-ph]} \BibitemShut {NoStop}%
\bibitem [{\citenamefont {Celi}\ \emph {et~al.}(2016)\citenamefont {Celi},
  \citenamefont {Sanpera}, \citenamefont {Ahufinger},\ and\ \citenamefont
  {Lewenstein}}]{Celi_2016}%
  \BibitemOpen
  \bibfield  {author} {\bibinfo {author} {\bibfnamefont {Alessio}\ \bibnamefont
  {Celi}}, \bibinfo {author} {\bibfnamefont {Anna}\ \bibnamefont {Sanpera}},
  \bibinfo {author} {\bibfnamefont {Veronica}\ \bibnamefont {Ahufinger}}, \
  and\ \bibinfo {author} {\bibfnamefont {Maciej}\ \bibnamefont {Lewenstein}},\
  }\bibfield  {title} {\enquote {\bibinfo {title} {Quantum optics and frontiers
  of physics: the third quantum revolution},}\ }\href {\doibase
  10.1088/1402-4896/92/1/013003} {\bibfield  {journal} {\bibinfo  {journal}
  {Physica Scripta}\ }\textbf {\bibinfo {volume} {92}},\ \bibinfo {pages}
  {013003} (\bibinfo {year} {2016})}\BibitemShut {NoStop}%
\bibitem [{\citenamefont {Gu}\ \emph {et~al.}(2016)\citenamefont {Gu},
  \citenamefont {Wang},\ and\ \citenamefont {Wen}}]{PhysRevB.93.115136}%
  \BibitemOpen
  \bibfield  {author} {\bibinfo {author} {\bibfnamefont {Zheng-Cheng}\
  \bibnamefont {Gu}}, \bibinfo {author} {\bibfnamefont {Juven~C.}\ \bibnamefont
  {Wang}}, \ and\ \bibinfo {author} {\bibfnamefont {Xiao-Gang}\ \bibnamefont
  {Wen}},\ }\bibfield  {title} {\enquote {\bibinfo {title} {Multikink
  topological terms and charge-binding domain-wall condensation induced
  symmetry-protected topological states: Beyond chern-simons/bf field
  theories},}\ }\href {\doibase 10.1103/PhysRevB.93.115136} {\bibfield
  {journal} {\bibinfo  {journal} {Phys. Rev. B}\ }\textbf {\bibinfo {volume}
  {93}},\ \bibinfo {pages} {115136} (\bibinfo {year} {2016})}\BibitemShut
  {NoStop}%
\bibitem [{\citenamefont {Ye}\ and\ \citenamefont {Gu}(2015)}]{bti2}%
  \BibitemOpen
  \bibfield  {author} {\bibinfo {author} {\bibfnamefont {Peng}\ \bibnamefont
  {Ye}}\ and\ \bibinfo {author} {\bibfnamefont {Zheng-Cheng}\ \bibnamefont
  {Gu}},\ }\bibfield  {title} {\enquote {\bibinfo {title} {Vortex-line
  condensation in three dimensions: A physical mechanism for bosonic
  topological insulators},}\ }\href {\doibase 10.1103/PhysRevX.5.021029}
  {\bibfield  {journal} {\bibinfo  {journal} {Phys. Rev. X}\ }\textbf {\bibinfo
  {volume} {5}},\ \bibinfo {pages} {021029} (\bibinfo {year}
  {2015})}\BibitemShut {NoStop}%
\bibitem [{\citenamefont {Ye}\ and\ \citenamefont {Gu}(2016)}]{YeGu2015}%
  \BibitemOpen
  \bibfield  {author} {\bibinfo {author} {\bibfnamefont {Peng}\ \bibnamefont
  {Ye}}\ and\ \bibinfo {author} {\bibfnamefont {Zheng-Cheng}\ \bibnamefont
  {Gu}},\ }\bibfield  {title} {\enquote {\bibinfo {title} {Topological quantum
  field theory of three-dimensional bosonic abelian-symmetry-protected
  topological phases},}\ }\href {\doibase 10.1103/PhysRevB.93.205157}
  {\bibfield  {journal} {\bibinfo  {journal} {Phys. Rev. B}\ }\textbf {\bibinfo
  {volume} {93}},\ \bibinfo {pages} {205157} (\bibinfo {year}
  {2016})}\BibitemShut {NoStop}%
\bibitem [{\citenamefont {Chan}\ \emph {et~al.}(2018)\citenamefont {Chan},
  \citenamefont {Ye},\ and\ \citenamefont {Ryu}}]{yp18prl}%
  \BibitemOpen
  \bibfield  {author} {\bibinfo {author} {\bibfnamefont {AtMa P.~O.}\
  \bibnamefont {Chan}}, \bibinfo {author} {\bibfnamefont {Peng}\ \bibnamefont
  {Ye}}, \ and\ \bibinfo {author} {\bibfnamefont {Shinsei}\ \bibnamefont
  {Ryu}},\ }\bibfield  {title} {\enquote {\bibinfo {title} {Braiding with
  borromean rings in ($3+1$)-dimensional spacetime},}\ }\href {\doibase
  10.1103/PhysRevLett.121.061601} {\bibfield  {journal} {\bibinfo  {journal}
  {Phys. Rev. Lett.}\ }\textbf {\bibinfo {volume} {121}},\ \bibinfo {pages}
  {061601} (\bibinfo {year} {2018})}\BibitemShut {NoStop}%
\bibitem [{\citenamefont {Wen}\ and\ \citenamefont {Zee}(1990)}]{WenSuperf}%
  \BibitemOpen
  \bibfield  {author} {\bibinfo {author} {\bibfnamefont {X.~G.}\ \bibnamefont
  {Wen}}\ and\ \bibinfo {author} {\bibfnamefont {A.}~\bibnamefont {Zee}},\
  }\bibfield  {title} {\enquote {\bibinfo {title} {Compressibility and
  superfluidity in the fractional-statistics liquid},}\ }\href {\doibase
  10.1103/PhysRevB.41.240} {\bibfield  {journal} {\bibinfo  {journal} {Phys.
  Rev. B}\ }\textbf {\bibinfo {volume} {41}},\ \bibinfo {pages} {240--253}
  (\bibinfo {year} {1990})}\BibitemShut {NoStop}%
\bibitem [{\citenamefont {Chen}\ \emph {et~al.}(2014)\citenamefont {Chen},
  \citenamefont {Lu},\ and\ \citenamefont {Vishwanath}}]{Chen:2014aa}%
  \BibitemOpen
  \bibfield  {author} {\bibinfo {author} {\bibfnamefont {Xie}\ \bibnamefont
  {Chen}}, \bibinfo {author} {\bibfnamefont {Yuan-Ming}\ \bibnamefont {Lu}}, \
  and\ \bibinfo {author} {\bibfnamefont {Ashvin}\ \bibnamefont {Vishwanath}},\
  }\bibfield  {title} {\enquote {\bibinfo {title} {Symmetry-protected
  topological phases from decorated domain walls},}\ }\href
  {http://dx.doi.org/10.1038/ncomms4507} {\bibfield  {journal} {\bibinfo
  {journal} {Nature Communications}\ }\textbf {\bibinfo {volume} {5}},\
  \bibinfo {pages} {3507 EP --} (\bibinfo {year} {2014})}\BibitemShut {NoStop}%
\bibitem [{\citenamefont {Chamon}(2005)}]{Chamon05}%
  \BibitemOpen
  \bibfield  {author} {\bibinfo {author} {\bibfnamefont {Claudio}\ \bibnamefont
  {Chamon}},\ }\bibfield  {title} {\enquote {\bibinfo {title} {Quantum
  glassiness in strongly correlated clean systems: An example of topological
  overprotection},}\ }\href {\doibase 10.1103/PhysRevLett.94.040402} {\bibfield
   {journal} {\bibinfo  {journal} {Phys. Rev. Lett.}\ }\textbf {\bibinfo
  {volume} {94}},\ \bibinfo {pages} {040402} (\bibinfo {year}
  {2005})}\BibitemShut {NoStop}%
\bibitem [{\citenamefont {Vijay}\ \emph {et~al.}(2015)\citenamefont {Vijay},
  \citenamefont {Haah},\ and\ \citenamefont {Fu}}]{Vijay2015}%
  \BibitemOpen
  \bibfield  {author} {\bibinfo {author} {\bibfnamefont {Sagar}\ \bibnamefont
  {Vijay}}, \bibinfo {author} {\bibfnamefont {Jeongwan}\ \bibnamefont {Haah}},
  \ and\ \bibinfo {author} {\bibfnamefont {Liang}\ \bibnamefont {Fu}},\
  }\bibfield  {title} {\enquote {\bibinfo {title} {A new kind of topological
  quantum order: A dimensional hierarchy of quasiparticles built from
  stationary excitations},}\ }\href {\doibase 10.1103/PhysRevB.92.235136}
  {\bibfield  {journal} {\bibinfo  {journal} {Phys. Rev. B}\ }\textbf {\bibinfo
  {volume} {92}},\ \bibinfo {pages} {235136} (\bibinfo {year}
  {2015})}\BibitemShut {NoStop}%
\bibitem [{\citenamefont {Vijay}\ \emph
  {et~al.}(2016{\natexlab{a}})\citenamefont {Vijay}, \citenamefont {Haah},\
  and\ \citenamefont {Fu}}]{Vijay2016a}%
  \BibitemOpen
  \bibfield  {author} {\bibinfo {author} {\bibfnamefont {Sagar}\ \bibnamefont
  {Vijay}}, \bibinfo {author} {\bibfnamefont {Jeongwan}\ \bibnamefont {Haah}},
  \ and\ \bibinfo {author} {\bibfnamefont {Liang}\ \bibnamefont {Fu}},\
  }\bibfield  {title} {\enquote {\bibinfo {title} {Fracton topological order,
  generalized lattice gauge theory, and duality},}\ }\href {\doibase
  10.1103/PhysRevB.94.235157} {\bibfield  {journal} {\bibinfo  {journal} {Phys.
  Rev. B}\ }\textbf {\bibinfo {volume} {94}},\ \bibinfo {pages} {235157}
  (\bibinfo {year} {2016}{\natexlab{a}})}\BibitemShut {NoStop}%
\bibitem [{\citenamefont {Nandkishore}\ and\ \citenamefont
  {Hermele}(2019)}]{fracton_review1}%
  \BibitemOpen
  \bibfield  {author} {\bibinfo {author} {\bibfnamefont {Rahul~M.}\
  \bibnamefont {Nandkishore}}\ and\ \bibinfo {author} {\bibfnamefont {Michael}\
  \bibnamefont {Hermele}},\ }\bibfield  {title} {\enquote {\bibinfo {title}
  {Fractons},}\ }\href {\doibase 10.1146/annurev-conmatphys-031218-013604}
  {\bibfield  {journal} {\bibinfo  {journal} {Annual Review of Condensed Matter
  Physics}\ }\textbf {\bibinfo {volume} {10}},\ \bibinfo {pages} {295--313}
  (\bibinfo {year} {2019})},\ \Eprint
  {http://arxiv.org/abs/https://doi.org/10.1146/annurev-conmatphys-031218-013604}
  {https://doi.org/10.1146/annurev-conmatphys-031218-013604} \BibitemShut
  {NoStop}%
\bibitem [{\citenamefont {Vijay}\ \emph
  {et~al.}(2016{\natexlab{b}})\citenamefont {Vijay}, \citenamefont {Haah},\
  and\ \citenamefont {Fu}}]{Vijay2016}%
  \BibitemOpen
  \bibfield  {author} {\bibinfo {author} {\bibfnamefont {Sagar}\ \bibnamefont
  {Vijay}}, \bibinfo {author} {\bibfnamefont {Jeongwan}\ \bibnamefont {Haah}},
  \ and\ \bibinfo {author} {\bibfnamefont {Liang}\ \bibnamefont {Fu}},\
  }\bibfield  {title} {\enquote {\bibinfo {title} {Fracton topological order,
  generalized lattice gauge theory, and duality},}\ }\href {\doibase
  10.1103/PhysRevB.94.235157} {\bibfield  {journal} {\bibinfo  {journal} {Phys.
  Rev. B}\ }\textbf {\bibinfo {volume} {94}},\ \bibinfo {pages} {235157}
  (\bibinfo {year} {2016}{\natexlab{b}})}\BibitemShut {NoStop}%
\bibitem [{\citenamefont {Prem}\ \emph {et~al.}(2017)\citenamefont {Prem},
  \citenamefont {Haah},\ and\ \citenamefont {Nandkishore}}]{Prem2017}%
  \BibitemOpen
  \bibfield  {author} {\bibinfo {author} {\bibfnamefont {Abhinav}\ \bibnamefont
  {Prem}}, \bibinfo {author} {\bibfnamefont {Jeongwan}\ \bibnamefont {Haah}}, \
  and\ \bibinfo {author} {\bibfnamefont {Rahul}\ \bibnamefont {Nandkishore}},\
  }\bibfield  {title} {\enquote {\bibinfo {title} {Glassy quantum dynamics in
  translation invariant fracton models},}\ }\href {\doibase
  10.1103/PhysRevB.95.155133} {\bibfield  {journal} {\bibinfo  {journal} {Phys.
  Rev. B}\ }\textbf {\bibinfo {volume} {95}},\ \bibinfo {pages} {155133}
  (\bibinfo {year} {2017})}\BibitemShut {NoStop}%
\bibitem [{\citenamefont {Shirley}\ \emph
  {et~al.}(2019{\natexlab{a}})\citenamefont {Shirley}, \citenamefont {Slagle},\
  and\ \citenamefont {Chen}}]{Shirley2019}%
  \BibitemOpen
  \bibfield  {author} {\bibinfo {author} {\bibfnamefont {Wilbur}\ \bibnamefont
  {Shirley}}, \bibinfo {author} {\bibfnamefont {Kevin}\ \bibnamefont {Slagle}},
  \ and\ \bibinfo {author} {\bibfnamefont {Xie}\ \bibnamefont {Chen}},\
  }\bibfield  {title} {\enquote {\bibinfo {title} {{Foliated fracton order from
  gauging subsystem symmetries}},}\ }\href {\doibase
  10.21468/SciPostPhys.6.4.041} {\bibfield  {journal} {\bibinfo  {journal}
  {SciPost Phys.}\ }\textbf {\bibinfo {volume} {6}},\ \bibinfo {pages} {41}
  (\bibinfo {year} {2019}{\natexlab{a}})}\BibitemShut {NoStop}%
\bibitem [{\citenamefont {Ma}\ \emph {et~al.}(2017)\citenamefont {Ma},
  \citenamefont {Lake}, \citenamefont {Chen},\ and\ \citenamefont
  {Hermele}}]{Ma2017}%
  \BibitemOpen
  \bibfield  {author} {\bibinfo {author} {\bibfnamefont {Han}\ \bibnamefont
  {Ma}}, \bibinfo {author} {\bibfnamefont {Ethan}\ \bibnamefont {Lake}},
  \bibinfo {author} {\bibfnamefont {Xie}\ \bibnamefont {Chen}}, \ and\ \bibinfo
  {author} {\bibfnamefont {Michael}\ \bibnamefont {Hermele}},\ }\bibfield
  {title} {\enquote {\bibinfo {title} {Fracton topological order via coupled
  layers},}\ }\href {\doibase 10.1103/PhysRevB.95.245126} {\bibfield  {journal}
  {\bibinfo  {journal} {Phys. Rev. B}\ }\textbf {\bibinfo {volume} {95}},\
  \bibinfo {pages} {245126} (\bibinfo {year} {2017})}\BibitemShut {NoStop}%
\bibitem [{\citenamefont {Haah}(2011)}]{Haah2011}%
  \BibitemOpen
  \bibfield  {author} {\bibinfo {author} {\bibfnamefont {Jeongwan}\
  \bibnamefont {Haah}},\ }\bibfield  {title} {\enquote {\bibinfo {title} {Local
  stabilizer codes in three dimensions without string logical operators},}\
  }\href {\doibase 10.1103/PhysRevA.83.042330} {\bibfield  {journal} {\bibinfo
  {journal} {Phys. Rev. A}\ }\textbf {\bibinfo {volume} {83}},\ \bibinfo
  {pages} {042330} (\bibinfo {year} {2011})}\BibitemShut {NoStop}%
\bibitem [{\citenamefont {{Bulmash}}\ and\ \citenamefont
  {{Barkeshli}}(2019)}]{Bulmash2019}%
  \BibitemOpen
  \bibfield  {author} {\bibinfo {author} {\bibfnamefont {Daniel}\ \bibnamefont
  {{Bulmash}}}\ and\ \bibinfo {author} {\bibfnamefont {Maissam}\ \bibnamefont
  {{Barkeshli}}},\ }\bibfield  {title} {\enquote {\bibinfo {title} {{Gauging
  fractons: immobile non-Abelian quasiparticles, fractals, and
  position-dependent degeneracies}},}\ }\href@noop {} {\bibfield  {journal}
  {\bibinfo  {journal} {arXiv e-prints}\ ,\ \bibinfo {eid} {arXiv:1905.05771}}
  (\bibinfo {year} {2019})},\ \Eprint {http://arxiv.org/abs/1905.05771}
  {arXiv:1905.05771 [cond-mat.str-el]} \BibitemShut {NoStop}%
\bibitem [{\citenamefont {{Prem}}\ and\ \citenamefont
  {{Williamson}}(2019)}]{Prem2019}%
  \BibitemOpen
  \bibfield  {author} {\bibinfo {author} {\bibfnamefont {A.}~\bibnamefont
  {{Prem}}}\ and\ \bibinfo {author} {\bibfnamefont {D.~J.}\ \bibnamefont
  {{Williamson}}},\ }\bibfield  {title} {\enquote {\bibinfo {title} {{Gauging
  permutation symmetries as a route to non-Abelian fractons}},}\ }\href@noop {}
  {\bibfield  {journal} {\bibinfo  {journal} {arXiv e-prints}\ } (\bibinfo
  {year} {2019})},\ \Eprint {http://arxiv.org/abs/1905.06309} {arXiv:1905.06309
  [cond-mat.str-el]} \BibitemShut {NoStop}%
\bibitem [{\citenamefont {{Bulmash}}\ and\ \citenamefont
  {{Barkeshli}}(2018)}]{Bulmash2018}%
  \BibitemOpen
  \bibfield  {author} {\bibinfo {author} {\bibfnamefont {D.}~\bibnamefont
  {{Bulmash}}}\ and\ \bibinfo {author} {\bibfnamefont {M.}~\bibnamefont
  {{Barkeshli}}},\ }\bibfield  {title} {\enquote {\bibinfo {title}
  {{Generalized $U(1)$ Gauge Field Theories and Fractal Dynamics}},}\
  }\href@noop {} {\bibfield  {journal} {\bibinfo  {journal} {arXiv e-prints}\ }
  (\bibinfo {year} {2018})},\ \Eprint {http://arxiv.org/abs/1806.01855}
  {arXiv:1806.01855 [cond-mat.str-el]} \BibitemShut {NoStop}%
\bibitem [{\citenamefont {{Tian}}\ \emph {et~al.}(2018)\citenamefont {{Tian}},
  \citenamefont {{Samperton}},\ and\ \citenamefont {{Wang}}}]{Tian2018}%
  \BibitemOpen
  \bibfield  {author} {\bibinfo {author} {\bibfnamefont {K.~T.}\ \bibnamefont
  {{Tian}}}, \bibinfo {author} {\bibfnamefont {E.}~\bibnamefont {{Samperton}}},
  \ and\ \bibinfo {author} {\bibfnamefont {Z.}~\bibnamefont {{Wang}}},\
  }\bibfield  {title} {\enquote {\bibinfo {title} {{Haah codes on general three
  manifolds}},}\ }\href@noop {} {\bibfield  {journal} {\bibinfo  {journal}
  {arXiv e-prints}\ } (\bibinfo {year} {2018})},\ \Eprint
  {http://arxiv.org/abs/1812.02101} {arXiv:1812.02101 [quant-ph]} \BibitemShut
  {NoStop}%
\bibitem [{\citenamefont {{You}}\ \emph {et~al.}(2018)\citenamefont {{You}},
  \citenamefont {{Litinski}},\ and\ \citenamefont {{von Oppen}}}]{You2018}%
  \BibitemOpen
  \bibfield  {author} {\bibinfo {author} {\bibfnamefont {Yizhi}\ \bibnamefont
  {{You}}}, \bibinfo {author} {\bibfnamefont {Daniel}\ \bibnamefont
  {{Litinski}}}, \ and\ \bibinfo {author} {\bibfnamefont {Felix}\ \bibnamefont
  {{von Oppen}}},\ }\bibfield  {title} {\enquote {\bibinfo {title} {{Higher
  order topological superconductors as generators of quantum codes}},}\
  }\href@noop {} {\bibfield  {journal} {\bibinfo  {journal} {arXiv e-prints}\
  ,\ \bibinfo {eid} {arXiv:1810.10556}} (\bibinfo {year} {2018})},\ \Eprint
  {http://arxiv.org/abs/1810.10556} {arXiv:1810.10556 [cond-mat.str-el]}
  \BibitemShut {NoStop}%
\bibitem [{\citenamefont {Ma}\ \emph {et~al.}(2018)\citenamefont {Ma},
  \citenamefont {Hermele},\ and\ \citenamefont {Chen}}]{Ma2018}%
  \BibitemOpen
  \bibfield  {author} {\bibinfo {author} {\bibfnamefont {Han}\ \bibnamefont
  {Ma}}, \bibinfo {author} {\bibfnamefont {Michael}\ \bibnamefont {Hermele}}, \
  and\ \bibinfo {author} {\bibfnamefont {Xie}\ \bibnamefont {Chen}},\
  }\bibfield  {title} {\enquote {\bibinfo {title} {Fracton topological order
  from the higgs and partial-confinement mechanisms of rank-two gauge
  theory},}\ }\href {\doibase 10.1103/PhysRevB.98.035111} {\bibfield  {journal}
  {\bibinfo  {journal} {Phys. Rev. B}\ }\textbf {\bibinfo {volume} {98}},\
  \bibinfo {pages} {035111} (\bibinfo {year} {2018})}\BibitemShut {NoStop}%
\bibitem [{\citenamefont {Slagle}\ and\ \citenamefont
  {Kim}(2017)}]{Slagle2017}%
  \BibitemOpen
  \bibfield  {author} {\bibinfo {author} {\bibfnamefont {Kevin}\ \bibnamefont
  {Slagle}}\ and\ \bibinfo {author} {\bibfnamefont {Yong~Baek}\ \bibnamefont
  {Kim}},\ }\bibfield  {title} {\enquote {\bibinfo {title} {Fracton topological
  order from nearest-neighbor two-spin interactions and dualities},}\ }\href
  {\doibase 10.1103/PhysRevB.96.165106} {\bibfield  {journal} {\bibinfo
  {journal} {Phys. Rev. B}\ }\textbf {\bibinfo {volume} {96}},\ \bibinfo
  {pages} {165106} (\bibinfo {year} {2017})}\BibitemShut {NoStop}%
\bibitem [{\citenamefont {Hal\'asz}\ \emph {et~al.}(2017)\citenamefont
  {Hal\'asz}, \citenamefont {Hsieh},\ and\ \citenamefont
  {Balents}}]{Halasz2017}%
  \BibitemOpen
  \bibfield  {author} {\bibinfo {author} {\bibfnamefont {G\'abor~B.}\
  \bibnamefont {Hal\'asz}}, \bibinfo {author} {\bibfnamefont {Timothy~H.}\
  \bibnamefont {Hsieh}}, \ and\ \bibinfo {author} {\bibfnamefont {Leon}\
  \bibnamefont {Balents}},\ }\bibfield  {title} {\enquote {\bibinfo {title}
  {Fracton topological phases from strongly coupled spin chains},}\ }\href
  {\doibase 10.1103/PhysRevLett.119.257202} {\bibfield  {journal} {\bibinfo
  {journal} {Phys. Rev. Lett.}\ }\textbf {\bibinfo {volume} {119}},\ \bibinfo
  {pages} {257202} (\bibinfo {year} {2017})}\BibitemShut {NoStop}%
\bibitem [{\citenamefont {{Tian}}\ and\ \citenamefont
  {{Wang}}(2019)}]{Tian2019}%
  \BibitemOpen
  \bibfield  {author} {\bibinfo {author} {\bibfnamefont {Kevin~T.}\
  \bibnamefont {{Tian}}}\ and\ \bibinfo {author} {\bibfnamefont {Zhenghan}\
  \bibnamefont {{Wang}}},\ }\bibfield  {title} {\enquote {\bibinfo {title}
  {{Generalized Haah Codes and Fracton Models}},}\ }\href@noop {} {\bibfield
  {journal} {\bibinfo  {journal} {arXiv e-prints}\ ,\ \bibinfo {eid}
  {arXiv:1902.04543}} (\bibinfo {year} {2019})},\ \Eprint
  {http://arxiv.org/abs/1902.04543} {arXiv:1902.04543 [quant-ph]} \BibitemShut
  {NoStop}%
\bibitem [{\citenamefont {Shirley}\ \emph
  {et~al.}(2019{\natexlab{b}})\citenamefont {Shirley}, \citenamefont {Slagle},\
  and\ \citenamefont {Chen}}]{Shirley2019b}%
  \BibitemOpen
  \bibfield  {author} {\bibinfo {author} {\bibfnamefont {Wilbur}\ \bibnamefont
  {Shirley}}, \bibinfo {author} {\bibfnamefont {Kevin}\ \bibnamefont {Slagle}},
  \ and\ \bibinfo {author} {\bibfnamefont {Xie}\ \bibnamefont {Chen}},\
  }\bibfield  {title} {\enquote {\bibinfo {title} {{Foliated fracton order from
  gauging subsystem symmetries}},}\ }\href {\doibase
  10.21468/SciPostPhys.6.4.041} {\bibfield  {journal} {\bibinfo  {journal}
  {SciPost Phys.}\ }\textbf {\bibinfo {volume} {6}},\ \bibinfo {pages} {41}
  (\bibinfo {year} {2019}{\natexlab{b}})}\BibitemShut {NoStop}%
\bibitem [{\citenamefont {{Shirley}}\ \emph {et~al.}(2018)\citenamefont
  {{Shirley}}, \citenamefont {{Slagle}},\ and\ \citenamefont
  {{Chen}}}]{Shirley2018a}%
  \BibitemOpen
  \bibfield  {author} {\bibinfo {author} {\bibfnamefont {W.}~\bibnamefont
  {{Shirley}}}, \bibinfo {author} {\bibfnamefont {K.}~\bibnamefont {{Slagle}}},
  \ and\ \bibinfo {author} {\bibfnamefont {X.}~\bibnamefont {{Chen}}},\
  }\bibfield  {title} {\enquote {\bibinfo {title} {{Fractional excitations in
  foliated fracton phases}},}\ }\href@noop {} {\bibfield  {journal} {\bibinfo
  {journal} {arXiv e-prints}\ } (\bibinfo {year} {2018})},\ \Eprint
  {http://arxiv.org/abs/1806.08625} {arXiv:1806.08625 [cond-mat.str-el]}
  \BibitemShut {NoStop}%
\bibitem [{\citenamefont {Slagle}\ \emph {et~al.}(2019)\citenamefont {Slagle},
  \citenamefont {Aasen},\ and\ \citenamefont {Williamson}}]{Slagle2019a}%
  \BibitemOpen
  \bibfield  {author} {\bibinfo {author} {\bibfnamefont {Kevin}\ \bibnamefont
  {Slagle}}, \bibinfo {author} {\bibfnamefont {David}\ \bibnamefont {Aasen}}, \
  and\ \bibinfo {author} {\bibfnamefont {Dominic}\ \bibnamefont {Williamson}},\
  }\bibfield  {title} {\enquote {\bibinfo {title} {{Foliated Field Theory and
  String-Membrane-Net Condensation Picture of Fracton Order}},}\ }\href
  {\doibase 10.21468/SciPostPhys.6.4.043} {\bibfield  {journal} {\bibinfo
  {journal} {SciPost Phys.}\ }\textbf {\bibinfo {volume} {6}},\ \bibinfo
  {pages} {43} (\bibinfo {year} {2019})}\BibitemShut {NoStop}%
\bibitem [{\citenamefont {Shirley}\ \emph {et~al.}(2018)\citenamefont
  {Shirley}, \citenamefont {Slagle}, \citenamefont {Wang},\ and\ \citenamefont
  {Chen}}]{Shirley2018}%
  \BibitemOpen
  \bibfield  {author} {\bibinfo {author} {\bibfnamefont {Wilbur}\ \bibnamefont
  {Shirley}}, \bibinfo {author} {\bibfnamefont {Kevin}\ \bibnamefont {Slagle}},
  \bibinfo {author} {\bibfnamefont {Zhenghan}\ \bibnamefont {Wang}}, \ and\
  \bibinfo {author} {\bibfnamefont {Xie}\ \bibnamefont {Chen}},\ }\bibfield
  {title} {\enquote {\bibinfo {title} {Fracton models on general
  three-dimensional manifolds},}\ }\href {\doibase 10.1103/PhysRevX.8.031051}
  {\bibfield  {journal} {\bibinfo  {journal} {Phys. Rev. X}\ }\textbf {\bibinfo
  {volume} {8}},\ \bibinfo {pages} {031051} (\bibinfo {year}
  {2018})}\BibitemShut {NoStop}%
\bibitem [{\citenamefont {Prem}\ \emph {et~al.}(2019)\citenamefont {Prem},
  \citenamefont {Huang}, \citenamefont {Song},\ and\ \citenamefont
  {Hermele}}]{Prem2018}%
  \BibitemOpen
  \bibfield  {author} {\bibinfo {author} {\bibfnamefont {Abhinav}\ \bibnamefont
  {Prem}}, \bibinfo {author} {\bibfnamefont {Sheng-Jie}\ \bibnamefont {Huang}},
  \bibinfo {author} {\bibfnamefont {Hao}\ \bibnamefont {Song}}, \ and\ \bibinfo
  {author} {\bibfnamefont {Michael}\ \bibnamefont {Hermele}},\ }\bibfield
  {title} {\enquote {\bibinfo {title} {Cage-net fracton models},}\ }\href
  {\doibase 10.1103/PhysRevX.9.021010} {\bibfield  {journal} {\bibinfo
  {journal} {Phys. Rev. X}\ }\textbf {\bibinfo {volume} {9}},\ \bibinfo {pages}
  {021010} (\bibinfo {year} {2019})}\BibitemShut {NoStop}%
\bibitem [{\citenamefont {Pai}\ \emph {et~al.}(2019)\citenamefont {Pai},
  \citenamefont {Pretko},\ and\ \citenamefont {Nandkishore}}]{Pai2019}%
  \BibitemOpen
  \bibfield  {author} {\bibinfo {author} {\bibfnamefont {Shriya}\ \bibnamefont
  {Pai}}, \bibinfo {author} {\bibfnamefont {Michael}\ \bibnamefont {Pretko}}, \
  and\ \bibinfo {author} {\bibfnamefont {Rahul~M.}\ \bibnamefont
  {Nandkishore}},\ }\bibfield  {title} {\enquote {\bibinfo {title}
  {Localization in fractonic random circuits},}\ }\href {\doibase
  10.1103/PhysRevX.9.021003} {\bibfield  {journal} {\bibinfo  {journal} {Phys.
  Rev. X}\ }\textbf {\bibinfo {volume} {9}},\ \bibinfo {pages} {021003}
  (\bibinfo {year} {2019})}\BibitemShut {NoStop}%
\bibitem [{\citenamefont {{Pai}}\ and\ \citenamefont
  {{Pretko}}(2019)}]{Pai2019a}%
  \BibitemOpen
  \bibfield  {author} {\bibinfo {author} {\bibfnamefont {Shriya}\ \bibnamefont
  {{Pai}}}\ and\ \bibinfo {author} {\bibfnamefont {Michael}\ \bibnamefont
  {{Pretko}}},\ }\bibfield  {title} {\enquote {\bibinfo {title} {{Dynamical
  scar states in driven fracton systems}},}\ }\href@noop {} {\bibfield
  {journal} {\bibinfo  {journal} {arXiv e-prints}\ ,\ \bibinfo {eid}
  {arXiv:1903.06173}} (\bibinfo {year} {2019})},\ \Eprint
  {http://arxiv.org/abs/1903.06173} {arXiv:1903.06173 [cond-mat.stat-mech]}
  \BibitemShut {NoStop}%
\bibitem [{\citenamefont {{Sala}}\ \emph {et~al.}(2019)\citenamefont {{Sala}},
  \citenamefont {{Rakovszky}}, \citenamefont {{Verresen}}, \citenamefont
  {{Knap}},\ and\ \citenamefont {{Pollmann}}}]{Sala2019}%
  \BibitemOpen
  \bibfield  {author} {\bibinfo {author} {\bibfnamefont {Pablo}\ \bibnamefont
  {{Sala}}}, \bibinfo {author} {\bibfnamefont {Tibor}\ \bibnamefont
  {{Rakovszky}}}, \bibinfo {author} {\bibfnamefont {Ruben}\ \bibnamefont
  {{Verresen}}}, \bibinfo {author} {\bibfnamefont {Michael}\ \bibnamefont
  {{Knap}}}, \ and\ \bibinfo {author} {\bibfnamefont {Frank}\ \bibnamefont
  {{Pollmann}}},\ }\bibfield  {title} {\enquote {\bibinfo {title}
  {{Ergodicity-breaking arising from Hilbert space fragmentation in
  dipole-conserving Hamiltonians}},}\ }\href@noop {} {\bibfield  {journal}
  {\bibinfo  {journal} {arXiv e-prints}\ ,\ \bibinfo {eid} {arXiv:1904.04266}}
  (\bibinfo {year} {2019})},\ \Eprint {http://arxiv.org/abs/1904.04266}
  {arXiv:1904.04266 [cond-mat.str-el]} \BibitemShut {NoStop}%
\bibitem [{\citenamefont {Kumar}\ and\ \citenamefont
  {Potter}(2019)}]{Kumar2018}%
  \BibitemOpen
  \bibfield  {author} {\bibinfo {author} {\bibfnamefont {Ajesh}\ \bibnamefont
  {Kumar}}\ and\ \bibinfo {author} {\bibfnamefont {Andrew~C.}\ \bibnamefont
  {Potter}},\ }\bibfield  {title} {\enquote {\bibinfo {title}
  {Symmetry-enforced fractonicity and two-dimensional quantum crystal
  melting},}\ }\href {\doibase 10.1103/PhysRevB.100.045119} {\bibfield
  {journal} {\bibinfo  {journal} {Phys. Rev. B}\ }\textbf {\bibinfo {volume}
  {100}},\ \bibinfo {pages} {045119} (\bibinfo {year} {2019})}\BibitemShut
  {NoStop}%
\bibitem [{\citenamefont {Pretko}(2018)}]{Pretko2018}%
  \BibitemOpen
  \bibfield  {author} {\bibinfo {author} {\bibfnamefont {Michael}\ \bibnamefont
  {Pretko}},\ }\bibfield  {title} {\enquote {\bibinfo {title} {The fracton
  gauge principle},}\ }\href {\doibase 10.1103/PhysRevB.98.115134} {\bibfield
  {journal} {\bibinfo  {journal} {Phys. Rev. B}\ }\textbf {\bibinfo {volume}
  {98}},\ \bibinfo {pages} {115134} (\bibinfo {year} {2018})}\BibitemShut
  {NoStop}%
\bibitem [{\citenamefont {Pretko}(2017{\natexlab{a}})}]{Pretko2017}%
  \BibitemOpen
  \bibfield  {author} {\bibinfo {author} {\bibfnamefont {Michael}\ \bibnamefont
  {Pretko}},\ }\bibfield  {title} {\enquote {\bibinfo {title} {Subdimensional
  particle structure of higher rank $u(1)$ spin liquids},}\ }\href {\doibase
  10.1103/PhysRevB.95.115139} {\bibfield  {journal} {\bibinfo  {journal} {Phys.
  Rev. B}\ }\textbf {\bibinfo {volume} {95}},\ \bibinfo {pages} {115139}
  (\bibinfo {year} {2017}{\natexlab{a}})}\BibitemShut {NoStop}%
\bibitem [{\citenamefont {{Li}}\ and\ \citenamefont {{Ye}}(2019)}]{ye19a}%
  \BibitemOpen
  \bibfield  {author} {\bibinfo {author} {\bibfnamefont {Meng-Yuan}\
  \bibnamefont {{Li}}}\ and\ \bibinfo {author} {\bibfnamefont {Peng}\
  \bibnamefont {{Ye}}},\ }\bibfield  {title} {\enquote {\bibinfo {title}
  {{Exactly Solvable Fracton Models for Spatially Extended Excitations}},}\
  }\href@noop {} {\bibfield  {journal} {\bibinfo  {journal} {arXiv e-prints}\
  ,\ \bibinfo {eid} {arXiv:1909.02814}} (\bibinfo {year} {2019})},\ \Eprint
  {http://arxiv.org/abs/1909.02814} {arXiv:1909.02814 [cond-mat.str-el]}
  \BibitemShut {NoStop}%
\bibitem [{\citenamefont {Pretko}(2017{\natexlab{b}})}]{Pretko2017a}%
  \BibitemOpen
  \bibfield  {author} {\bibinfo {author} {\bibfnamefont {Michael}\ \bibnamefont
  {Pretko}},\ }\bibfield  {title} {\enquote {\bibinfo {title} {Generalized
  electromagnetism of subdimensional particles: A spin liquid story},}\ }\href
  {\doibase 10.1103/PhysRevB.96.035119} {\bibfield  {journal} {\bibinfo
  {journal} {Phys. Rev. B}\ }\textbf {\bibinfo {volume} {96}},\ \bibinfo
  {pages} {035119} (\bibinfo {year} {2017}{\natexlab{b}})}\BibitemShut
  {NoStop}%
\bibitem [{\citenamefont {{Radzihovsky}}\ and\ \citenamefont
  {{Hermele}}(2019)}]{Radzihovsky2019}%
  \BibitemOpen
  \bibfield  {author} {\bibinfo {author} {\bibfnamefont {L.}~\bibnamefont
  {{Radzihovsky}}}\ and\ \bibinfo {author} {\bibfnamefont {M.}~\bibnamefont
  {{Hermele}}},\ }\bibfield  {title} {\enquote {\bibinfo {title} {{Fractons
  from vector gauge theory}},}\ }\href@noop {} {\bibfield  {journal} {\bibinfo
  {journal} {arXiv e-prints}\ } (\bibinfo {year} {2019})},\ \Eprint
  {http://arxiv.org/abs/1905.06951} {arXiv:1905.06951 [cond-mat.str-el]}
  \BibitemShut {NoStop}%
\bibitem [{\citenamefont {{Dua}}\ \emph {et~al.}(2019)\citenamefont {{Dua}},
  \citenamefont {{Kim}}, \citenamefont {{Cheng}},\ and\ \citenamefont
  {{Williamson}}}]{Dua2019}%
  \BibitemOpen
  \bibfield  {author} {\bibinfo {author} {\bibfnamefont {A.}~\bibnamefont
  {{Dua}}}, \bibinfo {author} {\bibfnamefont {I.~H.}\ \bibnamefont {{Kim}}},
  \bibinfo {author} {\bibfnamefont {M.}~\bibnamefont {{Cheng}}}, \ and\
  \bibinfo {author} {\bibfnamefont {D.~J.}\ \bibnamefont {{Williamson}}},\
  }\bibfield  {title} {\enquote {\bibinfo {title} {{Sorting topological
  stabilizer models in three dimensions}},}\ }\href@noop {} {\bibfield
  {journal} {\bibinfo  {journal} {arXiv e-prints}\ } (\bibinfo {year}
  {2019})},\ \Eprint {http://arxiv.org/abs/1908.08049} {arXiv:1908.08049
  [quant-ph]} \BibitemShut {NoStop}%
\bibitem [{\citenamefont
  {Gromov}(2019{\natexlab{a}})}]{PhysRevLett.122.076403}%
  \BibitemOpen
  \bibfield  {author} {\bibinfo {author} {\bibfnamefont {Andrey}\ \bibnamefont
  {Gromov}},\ }\bibfield  {title} {\enquote {\bibinfo {title} {Chiral
  topological elasticity and fracton order},}\ }\href {\doibase
  10.1103/PhysRevLett.122.076403} {\bibfield  {journal} {\bibinfo  {journal}
  {Phys. Rev. Lett.}\ }\textbf {\bibinfo {volume} {122}},\ \bibinfo {pages}
  {076403} (\bibinfo {year} {2019}{\natexlab{a}})}\BibitemShut {NoStop}%
\bibitem [{\citenamefont {{Haah}}(2013)}]{haahthesis}%
  \BibitemOpen
  \bibfield  {author} {\bibinfo {author} {\bibfnamefont {Jeongwan}\
  \bibnamefont {{Haah}}},\ }\emph {\bibinfo {title} {{Lattice quantum codes and
  exotic topological phases of matter}}},\ \href@noop {} {Ph.D. thesis},\
  \bibinfo  {school} {California Institute of Technology} (\bibinfo {year}
  {2013})\BibitemShut {NoStop}%
\bibitem [{\citenamefont {Gromov}(2019{\natexlab{b}})}]{PhysRevX.9.031035}%
  \BibitemOpen
  \bibfield  {author} {\bibinfo {author} {\bibfnamefont {Andrey}\ \bibnamefont
  {Gromov}},\ }\bibfield  {title} {\enquote {\bibinfo {title} {Towards
  classification of fracton phases: The multipole algebra},}\ }\href {\doibase
  10.1103/PhysRevX.9.031035} {\bibfield  {journal} {\bibinfo  {journal} {Phys.
  Rev. X}\ }\textbf {\bibinfo {volume} {9}},\ \bibinfo {pages} {031035}
  (\bibinfo {year} {2019}{\natexlab{b}})}\BibitemShut {NoStop}%
\bibitem [{\citenamefont {{You}}\ \emph {et~al.}(2019)\citenamefont {{You}},
  \citenamefont {{Devakul}}, \citenamefont {{Sondhi}},\ and\ \citenamefont
  {{Burnell}}}]{2019arXiv190411530Y}%
  \BibitemOpen
  \bibfield  {author} {\bibinfo {author} {\bibfnamefont {Yizhi}\ \bibnamefont
  {{You}}}, \bibinfo {author} {\bibfnamefont {Trithep}\ \bibnamefont
  {{Devakul}}}, \bibinfo {author} {\bibfnamefont {S.~L.}\ \bibnamefont
  {{Sondhi}}}, \ and\ \bibinfo {author} {\bibfnamefont {F.~J.}\ \bibnamefont
  {{Burnell}}},\ }\bibfield  {title} {\enquote {\bibinfo {title} {{Fractonic
  Chern-Simons and BF theories}},}\ }\href@noop {} {\bibfield  {journal}
  {\bibinfo  {journal} {arXiv e-prints}\ ,\ \bibinfo {eid} {arXiv:1904.11530}}
  (\bibinfo {year} {2019})},\ \Eprint {http://arxiv.org/abs/1904.11530}
  {arXiv:1904.11530 [cond-mat.str-el]} \BibitemShut {NoStop}%
\bibitem [{\citenamefont {{Sous}}\ and\ \citenamefont
  {{Pretko}}(2019)}]{2019arXiv190408424S}%
  \BibitemOpen
  \bibfield  {author} {\bibinfo {author} {\bibfnamefont {John}\ \bibnamefont
  {{Sous}}}\ and\ \bibinfo {author} {\bibfnamefont {Michael}\ \bibnamefont
  {{Pretko}}},\ }\bibfield  {title} {\enquote {\bibinfo {title} {{Fractons from
  polarons and hole-doped antiferromagnets: Microscopic realizations}},}\
  }\href@noop {} {\bibfield  {journal} {\bibinfo  {journal} {arXiv e-prints}\
  ,\ \bibinfo {eid} {arXiv:1904.08424}} (\bibinfo {year} {2019})},\ \Eprint
  {http://arxiv.org/abs/1904.08424} {arXiv:1904.08424 [cond-mat.str-el]}
  \BibitemShut {NoStop}%
\bibitem [{\citenamefont {{Wang}}\ and\ \citenamefont
  {{Xu}}(2019)}]{2019arXiv190913879W}%
  \BibitemOpen
  \bibfield  {author} {\bibinfo {author} {\bibfnamefont {Juven}\ \bibnamefont
  {{Wang}}}\ and\ \bibinfo {author} {\bibfnamefont {Kai}\ \bibnamefont
  {{Xu}}},\ }\bibfield  {title} {\enquote {\bibinfo {title} {{Higher-Rank
  Tensor Field Theory of Non-Abelian Fracton and Embeddon}},}\ }\href@noop {}
  {\bibfield  {journal} {\bibinfo  {journal} {arXiv e-prints}\ ,\ \bibinfo
  {eid} {arXiv:1909.13879}} (\bibinfo {year} {2019})},\ \Eprint
  {http://arxiv.org/abs/1909.13879} {arXiv:1909.13879 [hep-th]} \BibitemShut
  {NoStop}%
\bibitem [{\citenamefont {Pai}\ and\ \citenamefont
  {Pretko}(2018)}]{pretko18string}%
  \BibitemOpen
  \bibfield  {author} {\bibinfo {author} {\bibfnamefont {Shriya}\ \bibnamefont
  {Pai}}\ and\ \bibinfo {author} {\bibfnamefont {Michael}\ \bibnamefont
  {Pretko}},\ }\bibfield  {title} {\enquote {\bibinfo {title} {Fractonic line
  excitations: An inroad from three-dimensional elasticity theory},}\ }\href
  {\doibase 10.1103/PhysRevB.97.235102} {\bibfield  {journal} {\bibinfo
  {journal} {Phys. Rev. B}\ }\textbf {\bibinfo {volume} {97}},\ \bibinfo
  {pages} {235102} (\bibinfo {year} {2018})}\BibitemShut {NoStop}%
\bibitem [{\citenamefont {Pretko}\ and\ \citenamefont
  {Nandkishore}(2018)}]{pretko18localization}%
  \BibitemOpen
  \bibfield  {author} {\bibinfo {author} {\bibfnamefont {Michael}\ \bibnamefont
  {Pretko}}\ and\ \bibinfo {author} {\bibfnamefont {Rahul~M.}\ \bibnamefont
  {Nandkishore}},\ }\bibfield  {title} {\enquote {\bibinfo {title}
  {Localization of extended quantum objects},}\ }\href {\doibase
  10.1103/PhysRevB.98.134301} {\bibfield  {journal} {\bibinfo  {journal} {Phys.
  Rev. B}\ }\textbf {\bibinfo {volume} {98}},\ \bibinfo {pages} {134301}
  (\bibinfo {year} {2018})}\BibitemShut {NoStop}%
\bibitem [{\citenamefont {Williamson}\ \emph {et~al.}(2019)\citenamefont
  {Williamson}, \citenamefont {Bi},\ and\ \citenamefont
  {Cheng}}]{PhysRevB.100.125150}%
  \BibitemOpen
  \bibfield  {author} {\bibinfo {author} {\bibfnamefont {Dominic~J.}\
  \bibnamefont {Williamson}}, \bibinfo {author} {\bibfnamefont {Zhen}\
  \bibnamefont {Bi}}, \ and\ \bibinfo {author} {\bibfnamefont {Meng}\
  \bibnamefont {Cheng}},\ }\bibfield  {title} {\enquote {\bibinfo {title}
  {Fractonic matter in symmetry-enriched $u(1)$ gauge theory},}\ }\href
  {\doibase 10.1103/PhysRevB.100.125150} {\bibfield  {journal} {\bibinfo
  {journal} {Phys. Rev. B}\ }\textbf {\bibinfo {volume} {100}},\ \bibinfo
  {pages} {125150} (\bibinfo {year} {2019})}\BibitemShut {NoStop}%
\bibitem [{\citenamefont {Dua}\ \emph {et~al.}(2019)\citenamefont {Dua},
  \citenamefont {Williamson}, \citenamefont {Haah},\ and\ \citenamefont
  {Cheng}}]{PhysRevB.99.245135}%
  \BibitemOpen
  \bibfield  {author} {\bibinfo {author} {\bibfnamefont {Arpit}\ \bibnamefont
  {Dua}}, \bibinfo {author} {\bibfnamefont {Dominic~J.}\ \bibnamefont
  {Williamson}}, \bibinfo {author} {\bibfnamefont {Jeongwan}\ \bibnamefont
  {Haah}}, \ and\ \bibinfo {author} {\bibfnamefont {Meng}\ \bibnamefont
  {Cheng}},\ }\bibfield  {title} {\enquote {\bibinfo {title} {Compactifying
  fracton stabilizer models},}\ }\href {\doibase 10.1103/PhysRevB.99.245135}
  {\bibfield  {journal} {\bibinfo  {journal} {Phys. Rev. B}\ }\textbf {\bibinfo
  {volume} {99}},\ \bibinfo {pages} {245135} (\bibinfo {year}
  {2019})}\BibitemShut {NoStop}%
\bibitem [{\citenamefont {Shi}\ and\ \citenamefont
  {Lu}(2018)}]{PhysRevB.97.144106}%
  \BibitemOpen
  \bibfield  {author} {\bibinfo {author} {\bibfnamefont {Bowen}\ \bibnamefont
  {Shi}}\ and\ \bibinfo {author} {\bibfnamefont {Yuan-Ming}\ \bibnamefont
  {Lu}},\ }\bibfield  {title} {\enquote {\bibinfo {title} {Deciphering the
  nonlocal entanglement entropy of fracton topological orders},}\ }\href
  {\doibase 10.1103/PhysRevB.97.144106} {\bibfield  {journal} {\bibinfo
  {journal} {Phys. Rev. B}\ }\textbf {\bibinfo {volume} {97}},\ \bibinfo
  {pages} {144106} (\bibinfo {year} {2018})}\BibitemShut {NoStop}%
\bibitem [{\citenamefont {Song}\ \emph {et~al.}(2019)\citenamefont {Song},
  \citenamefont {Prem}, \citenamefont {Huang},\ and\ \citenamefont
  {Martin-Delgado}}]{PhysRevB.99.155118}%
  \BibitemOpen
  \bibfield  {author} {\bibinfo {author} {\bibfnamefont {Hao}\ \bibnamefont
  {Song}}, \bibinfo {author} {\bibfnamefont {Abhinav}\ \bibnamefont {Prem}},
  \bibinfo {author} {\bibfnamefont {Sheng-Jie}\ \bibnamefont {Huang}}, \ and\
  \bibinfo {author} {\bibfnamefont {M.~A.}\ \bibnamefont {Martin-Delgado}},\
  }\bibfield  {title} {\enquote {\bibinfo {title} {Twisted fracton models in
  three dimensions},}\ }\href {\doibase 10.1103/PhysRevB.99.155118} {\bibfield
  {journal} {\bibinfo  {journal} {Phys. Rev. B}\ }\textbf {\bibinfo {volume}
  {99}},\ \bibinfo {pages} {155118} (\bibinfo {year} {2019})}\BibitemShut
  {NoStop}%
\bibitem [{\citenamefont {Ma}\ and\ \citenamefont
  {Pretko}(2018)}]{MaHigherRankDQC}%
  \BibitemOpen
  \bibfield  {author} {\bibinfo {author} {\bibfnamefont {Han}\ \bibnamefont
  {Ma}}\ and\ \bibinfo {author} {\bibfnamefont {Michael}\ \bibnamefont
  {Pretko}},\ }\bibfield  {title} {\enquote {\bibinfo {title} {Higher-rank
  deconfined quantum criticality at the lifshitz transition and the exciton
  bose condensate},}\ }\href {\doibase 10.1103/PhysRevB.98.125105} {\bibfield
  {journal} {\bibinfo  {journal} {Phys. Rev. B}\ }\textbf {\bibinfo {volume}
  {98}},\ \bibinfo {pages} {125105} (\bibinfo {year} {2018})}\BibitemShut
  {NoStop}%
\bibitem [{\citenamefont {{Wang}}\ \emph {et~al.}(2019)\citenamefont {{Wang}},
  \citenamefont {{Xu}},\ and\ \citenamefont {{Yau}}}]{2019arXiv191101804W}%
  \BibitemOpen
  \bibfield  {author} {\bibinfo {author} {\bibfnamefont {Juven}\ \bibnamefont
  {{Wang}}}, \bibinfo {author} {\bibfnamefont {Kai}\ \bibnamefont {{Xu}}}, \
  and\ \bibinfo {author} {\bibfnamefont {Shing-Tung}\ \bibnamefont {{Yau}}},\
  }\bibfield  {title} {\enquote {\bibinfo {title} {{Higher-Rank Non-Abelian
  Tensor Field Theory: Higher-Moment or Subdimensional Polynomial Global
  Symmetry, Algebraic Variety, Noether's Theorem, and Gauge}},}\ }\href@noop {}
  {\bibfield  {journal} {\bibinfo  {journal} {arXiv e-prints}\ ,\ \bibinfo
  {eid} {arXiv:1911.01804}} (\bibinfo {year} {2019})},\ \Eprint
  {http://arxiv.org/abs/1911.01804} {arXiv:1911.01804 [hep-th]} \BibitemShut
  {NoStop}%
 \bibitem [{\citenamefont {{Doshi}}\ and\ \citenamefont
  {{Gromov}}(2020)}]{2020arXiv200503015D}%
  \BibitemOpen
  \bibfield  {author} {\bibinfo {author} {\bibfnamefont {D.}~\bibnamefont
  {{Doshi}}}\ and\ \bibinfo {author} {\bibfnamefont {A.}~\bibnamefont
  {{Gromov}}},\ }\href@noop {} {\bibfield  {journal} {\bibinfo  {journal}
  {arXiv e-prints}\ ,\ \bibinfo {eid} {arXiv:2005.03015}} (\bibinfo {year}
  {2020})},\ \Eprint {http://arxiv.org/abs/2005.03015} {arXiv:2005.03015
  [cond-mat.str-el]} \BibitemShut {NoStop}%
\bibitem [{\citenamefont {{Wang}}\ and\ \citenamefont
  {{Yau}}(2019)}]{2019arXiv191213485W}%
  \BibitemOpen
  \bibfield  {author} {\bibinfo {author} {\bibfnamefont {Juven}\ \bibnamefont
  {{Wang}}}\ and\ \bibinfo {author} {\bibfnamefont {Shing-Tung}\ \bibnamefont
  {{Yau}}},\ }\bibfield  {title} {\enquote {\bibinfo {title} {{Non-Abelian
  Gauged Fractonic Matter Field Theory: New Sigma Models, Superfluids and
  Vortices}},}\ }\href@noop {} {\bibfield  {journal} {\bibinfo  {journal}
  {arXiv e-prints}\ ,\ \bibinfo {eid} {arXiv:1912.13485}} (\bibinfo {year}
  {2019})},\ \Eprint {http://arxiv.org/abs/1912.13485} {arXiv:1912.13485
  [cond-mat.str-el]} \BibitemShut {NoStop}%
\bibitem [{\citenamefont {Ho{\v{r}}ava}(2009)}]{Hoava2009}%
  \BibitemOpen
  \bibfield  {author} {\bibinfo {author} {\bibfnamefont {Petr}\ \bibnamefont
  {Ho{\v{r}}ava}},\ }\bibfield  {title} {\enquote {\bibinfo {title} {Membranes
  at quantum criticality},}\ }\href {\doibase 10.1088/1126-6708/2009/03/020}
  {\bibfield  {journal} {\bibinfo  {journal} {Journal of High Energy Physics}\
  }\textbf {\bibinfo {volume} {2009}},\ \bibinfo {pages} {020--020} (\bibinfo
  {year} {2009})}\BibitemShut {NoStop}%
\bibitem [{\citenamefont {Ho\ifmmode~\check{r}\else
  \v{r}\fi{}ava}(2009)}]{Hoava2009a}%
  \BibitemOpen
  \bibfield  {author} {\bibinfo {author} {\bibfnamefont {Petr}\ \bibnamefont
  {Ho\ifmmode~\check{r}\else \v{r}\fi{}ava}},\ }\bibfield  {title} {\enquote
  {\bibinfo {title} {Quantum gravity at a lifshitz point},}\ }\href {\doibase
  10.1103/PhysRevD.79.084008} {\bibfield  {journal} {\bibinfo  {journal} {Phys.
  Rev. D}\ }\textbf {\bibinfo {volume} {79}},\ \bibinfo {pages} {084008}
  (\bibinfo {year} {2009})}\BibitemShut {NoStop}%
\bibitem [{\citenamefont {Xu}\ and\ \citenamefont {Ho\ifmmode~\check{r}\else
  \v{r}\fi{}ava}(2010)}]{Xu2010}%
  \BibitemOpen
  \bibfield  {author} {\bibinfo {author} {\bibfnamefont {Cenke}\ \bibnamefont
  {Xu}}\ and\ \bibinfo {author} {\bibfnamefont {Petr}\ \bibnamefont
  {Ho\ifmmode~\check{r}\else \v{r}\fi{}ava}},\ }\bibfield  {title} {\enquote
  {\bibinfo {title} {Emergent gravity at a lifshitz point from a bose liquid on
  the lattice},}\ }\href {\doibase 10.1103/PhysRevD.81.104033} {\bibfield
  {journal} {\bibinfo  {journal} {Phys. Rev. D}\ }\textbf {\bibinfo {volume}
  {81}},\ \bibinfo {pages} {104033} (\bibinfo {year} {2010})}\BibitemShut
  {NoStop}%
\bibitem [{\citenamefont {{B{\"u}chler}}\ \emph {et~al.}(2005)\citenamefont
  {{B{\"u}chler}}, \citenamefont {{Hermele}}, \citenamefont {{Huber}},
  \citenamefont {{Fisher}},\ and\ \citenamefont
  {{Zoller}}}]{ExperimentFisher2005}%
  \BibitemOpen
  \bibfield  {author} {\bibinfo {author} {\bibfnamefont {H.~P.}\ \bibnamefont
  {{B{\"u}chler}}}, \bibinfo {author} {\bibfnamefont {M.}~\bibnamefont
  {{Hermele}}}, \bibinfo {author} {\bibfnamefont {S.~D.}\ \bibnamefont
  {{Huber}}}, \bibinfo {author} {\bibfnamefont {Matthew~P.}\ \bibnamefont
  {{Fisher}}}, \ and\ \bibinfo {author} {\bibfnamefont {P.}~\bibnamefont
  {{Zoller}}},\ }\bibfield  {title} {\enquote {\bibinfo {title} {{Atomic
  Quantum Simulator for Lattice Gauge Theories and Ring Exchange Models}},}\
  }\href {\doibase 10.1103/PhysRevLett.95.040402} {\bibfield  {journal}
  {\bibinfo  {journal} {\prl}\ }\textbf {\bibinfo {volume} {95}},\ \bibinfo
  {eid} {040402} (\bibinfo {year} {2005})},\ \Eprint
  {http://arxiv.org/abs/cond-mat/0503254} {arXiv:cond-mat/0503254
  [cond-mat.str-el]} \BibitemShut {NoStop}%
\bibitem [{\citenamefont {{Paredes}}\ and\ \citenamefont
  {{Bloch}}(2008)}]{2008PhRvA77b3603P}%
  \BibitemOpen
  \bibfield  {author} {\bibinfo {author} {\bibfnamefont {Bel{\'e}n}\
  \bibnamefont {{Paredes}}}\ and\ \bibinfo {author} {\bibfnamefont {Immanuel}\
  \bibnamefont {{Bloch}}},\ }\bibfield  {title} {\enquote {\bibinfo {title}
  {{Minimum instances of topological matter in an optical plaquette}},}\ }\href
  {\doibase 10.1103/PhysRevA.77.023603} {\bibfield  {journal} {\bibinfo
  {journal} {\pra}\ }\textbf {\bibinfo {volume} {77}},\ \bibinfo {eid} {023603}
  (\bibinfo {year} {2008})},\ \Eprint {http://arxiv.org/abs/0711.3796}
  {arXiv:0711.3796 [cond-mat.other]} \BibitemShut {NoStop}%
\bibitem [{\citenamefont {{Seiberg}}(2019)}]{Seiberg2019arXiv1909}%
  \BibitemOpen
  \bibfield  {author} {\bibinfo {author} {\bibfnamefont {Nathan}\ \bibnamefont
  {{Seiberg}}},\ }\bibfield  {title} {\enquote {\bibinfo {title} {{Field
  Theories With a Vector Global Symmetry}},}\ }\href@noop {} {\bibfield
  {journal} {\bibinfo  {journal} {arXiv e-prints}\ ,\ \bibinfo {eid}
  {arXiv:1909.10544}} (\bibinfo {year} {2019})},\ \Eprint
  {http://arxiv.org/abs/1909.10544} {arXiv:1909.10544 [cond-mat.str-el]}
  \BibitemShut {NoStop}%
\bibitem [{\citenamefont {Paramekanti}\ \emph {et~al.}(2002)\citenamefont
  {Paramekanti}, \citenamefont {Balents},\ and\ \citenamefont
  {Fisher}}]{FisherEBL}%
  \BibitemOpen
  \bibfield  {author} {\bibinfo {author} {\bibfnamefont {Arun}\ \bibnamefont
  {Paramekanti}}, \bibinfo {author} {\bibfnamefont {Leon}\ \bibnamefont
  {Balents}}, \ and\ \bibinfo {author} {\bibfnamefont {Matthew P.~A.}\
  \bibnamefont {Fisher}},\ }\bibfield  {title} {\enquote {\bibinfo {title}
  {Ring exchange, the exciton bose liquid, and bosonization in two
  dimensions},}\ }\href {\doibase 10.1103/PhysRevB.66.054526} {\bibfield
  {journal} {\bibinfo  {journal} {Phys. Rev. B}\ }\textbf {\bibinfo {volume}
  {66}},\ \bibinfo {pages} {054526} (\bibinfo {year} {2002})}\BibitemShut
  {NoStop}%
\end{thebibliography}
 

\end{document}